\documentclass[12pt]{article}

\usepackage{mdwlist}
\usepackage{amsmath}
\usepackage{amsfonts}
\usepackage{amssymb}
\usepackage{graphicx}
\usepackage[margin=2cm,nohead]{geometry}
\usepackage{afterpage}
\usepackage{parskip}
\usepackage{authblk}
\usepackage{changes}
\usepackage{float}

\newcommand{\exval}[1]{\langle {#1} \rangle}
\newcommand{\var}[1]{\mathrm{var}({#1})}

\newcommand{\bi}{\begin{itemize}}
\newcommand{\ei}{\end{itemize}}
\newcommand{\be}{\begin{equation}}
\newcommand{\ee}{\end{equation}}
\newcommand{\id}{\mathbb{I}}
\newcommand{\comment}[1]{}

\newcommand{\mur}{\exval{r}}

\newcommand{\sigAll}{\sigma_{all}}

\begin{document}

\title{Interpretation of Correlated Neural Variability from Models of Feed-Forward and Recurrent Circuits}
\author{Volker Pernice $^{1,2}$ \and Rava Azeredo da Silveira $^{1,2,3,*}$}
\date{ \textit{$^1$ Department of Physics, \'Ecole Normale Sup\'erieure, PSL
Research University, 75005 Paris, France}\\
\textit{$^2$ Laboratoire de Physique Statistique, Centre National de la
Recherche Scientifique, Universit\'e Pierre et Marie Curie, Universit\'e
Denis Diderot, 75005 Paris, France}\\
\textit{$^3$ Princeton Neuroscience Institute, Princeton University,
Princeton, NJ 08544, USA}\\
\textit{$^*$ corresponding author}\\
\today
}

\maketitle{}

\abstract{
The correlated variability in the responses of a neural population to the repeated presentation of a sensory stimulus is a universally observed phenomenon. Such correlations have been studied in much detail, both with respect to their mechanistic origin and to their influence on stimulus discrimination and on the performance of population codes.  In particular, recurrent neural network models have been used to understand the origin (or lack) of correlations in neural activity. Here, we apply a model of recurrently connected stochastic neurons to interpret correlations found in a population of neurons recorded from mouse auditory cortex. We study the consequences of recurrent connections on the stimulus dependence of correlations, and we compare them to those from alternative sources of correlated variability, like correlated gain fluctuations and common input in feed-forward architectures. 

We find that a recurrent network model with random effective connections reproduces observed statistics, like the relation between noise and signal correlations in the data, in a natural way.  In the model, we can analyze directly the relation between network parameters, correlations, and how well pairs of stimuli can be discriminated based on population activity. In this way, we can relate circuit parameters to information processing. 
}

\section{Author Summary}

The response of neurons to a stimulus is variable across trials. A natural
solution for reliable coding in the face of noise is the averaging across a
neural population. The nature of this averaging depends on the structure of
noise correlations in the neural population. In turn, the correlation
structure depends on the way noise and correlations are generated in neural
circuits. It is in general difficult, however, to tease apart the origin of
correlations from the observed population activity alone. In this article,
we explore different theoretical scenarios of the way in which correlations
can be generated, and we relate these to the architecture of feed-forward
and recurrent neural circuits. Analyzing population recordings of the
activity in mouse auditory cortex in response to sound stimuli, we find that
population statistics are consistent with those generated in a recurrent
network model. Using this model, we can then quantify the effects of network
properties on average population responses, noise correlations, and the
representation of sensory information.

\section{Introduction}

In the search for clues about the function of neural circuits, it has become customary to rely upon recordings of the responses of large populations of neurons. These measurements exhibit the concerted activity of neural populations in different conditions, such as presentations of different stimuli, summarized by stimulus-dependent, high-dimensional statistics. 
The lowest moment of these statistics are also the best characterized, namely the mean response of neurons, the variability of response of single neurons, and their pairwise correlations. With these statistics in hand, one can ask two questions:
 How are they generated in the neural population?
 What purpose, if any, do they serve? While the first question is mechanistic and the second is functional, the two are intimately linked. We focus here on these questions and on their connection.

Along the mechanistic line of research, a number of network models have been
proposed to account for the statistics in measurements as well as to
determine the relationship between anatomical and physiological parameters,
on the one hand, and observed dynamics, on the other hand. 
 In this
spirit, for example models of balanced networks have been proposed to
explain asynchronous and irregular spike trains \cite{VanVreeswijk1998,Shadlen1998,Litwin-Kumar2012}.
Mechanistic explanations for the origin of pairwise correlation include the influence of recurrent connections \cite{Renart2010, Pernice2011, Trousdale2012} and global fluctuations (e.g., from top-down afferents) \cite{Goris2014,Scholvinck2015,Lin2015,Ecker2015,Doiron2016, Kohn2016}.
Reading the origin of correlations from the
recorded activity in a population of neurons is, however, a difficult task \cite{Pillow2008,Vidne2012,Goris2014,Ecker2014}. 

Along the functional line of research, the objective is to relate the
 statistics of neural responses to the function of neural circuits; for example, to
elucidate the role of the statistics in the representation of sensory
information. The relation between correlated variability in the population
response and the accuracy of stimulus representation, in particular, has
been the object of much study in recent years 
 \cite{Abbott1999,Wilke2002, Shamir2004,Averbeck2006a,Tkacik2010,Ecker2011,DaSilveira2014,Hu2014,Franke2015,Zylberberg2016}.
One reason for the focus on correlations is their possible effect in suppressing noise along relevant dimensions \cite{Zohary1994,Shadlen1998,Beck2011,Renart2012}.  Recent work has also illuminated the importance of the origin of correlations for their structure as it relates to the effect on stimulus representation \cite{Series2004,Moreno-bote2014,Kanitscheider2015a}. 

 In the present paper, we follow a three-pronged approach to
relate the possible mechanistic origins of correlation to recorded
statistics in cortical populations \cite{Bathellier2012}, on the one hand,
and to relate network mechanisms to the representation of information, on
the other hand. First, we study models of neural populations using the
framework of Poisson processes \cite{Hawkes1971} and we identify signatures in the population statistics
of different mechanistic motifs. Second, we
examine data on cortical populations in the light of our model results, and
we find that the measured statistics are consistent with those generated by
a recurrent network of Poisson neurons. Third, we use an idealized model of
a recurrent cortical population, in which we can manipulate mechanistic
parameters, to evaluate how the latter affect the accuracy of the
representation of information. 
We find that, if correlations are not too weak, correlations generated within recurrent networks can be distinguished from those that are due to external signals.  The estimated parameters from the data suggest that the interpretation of neural spike trains as Poisson processes implies a strong amplification together with noise generation by the  network dynamics.

\section{Materials and Methods}

\subsection{Experimental data set}
\label{sec:methodsExp}
The data set was first published and analyzed in \cite{Bathellier2012}. It
consists of the activity of neural populations (46-99 neurons) recorded
using calcium imaging in the auditory cortex of mice. Animals were isoflurane anesthetized (1 \%). Signals were obtained from neurons labeled  with the synthetic calcium indicator OGB1. Fluorescence was measured at 30 Hz sampling rate, and firing rates inferred from temporal deconvolution of the fluorescence signal.
 Up to 6
neural populations in each of 14 animals were recorded. The data points we
use are the average firing rates over a window of 250 ms after presentation
of each of 65 different sound stimuli, each for 15 trials. Responses were
measured relative to spontaneous activity. Hence, negative responses occurred, if the stimulus-evoked firing rates were smaller than the
spontaneous ones. The stimuli consisted of a range of different pure
tones 
 recorded for different sound intensities, as
well as a number of natural sounds. We do not distinguish individual
stimulus identities, but instead regard the set of stimuli as a diverse and,
to some degree, generic ensemble.

\subsection{Description of response variability}
\label{sec:methodsRepVar}
\begin{figure}[H]
\begin{center}
\includegraphics[width=13.2cm]{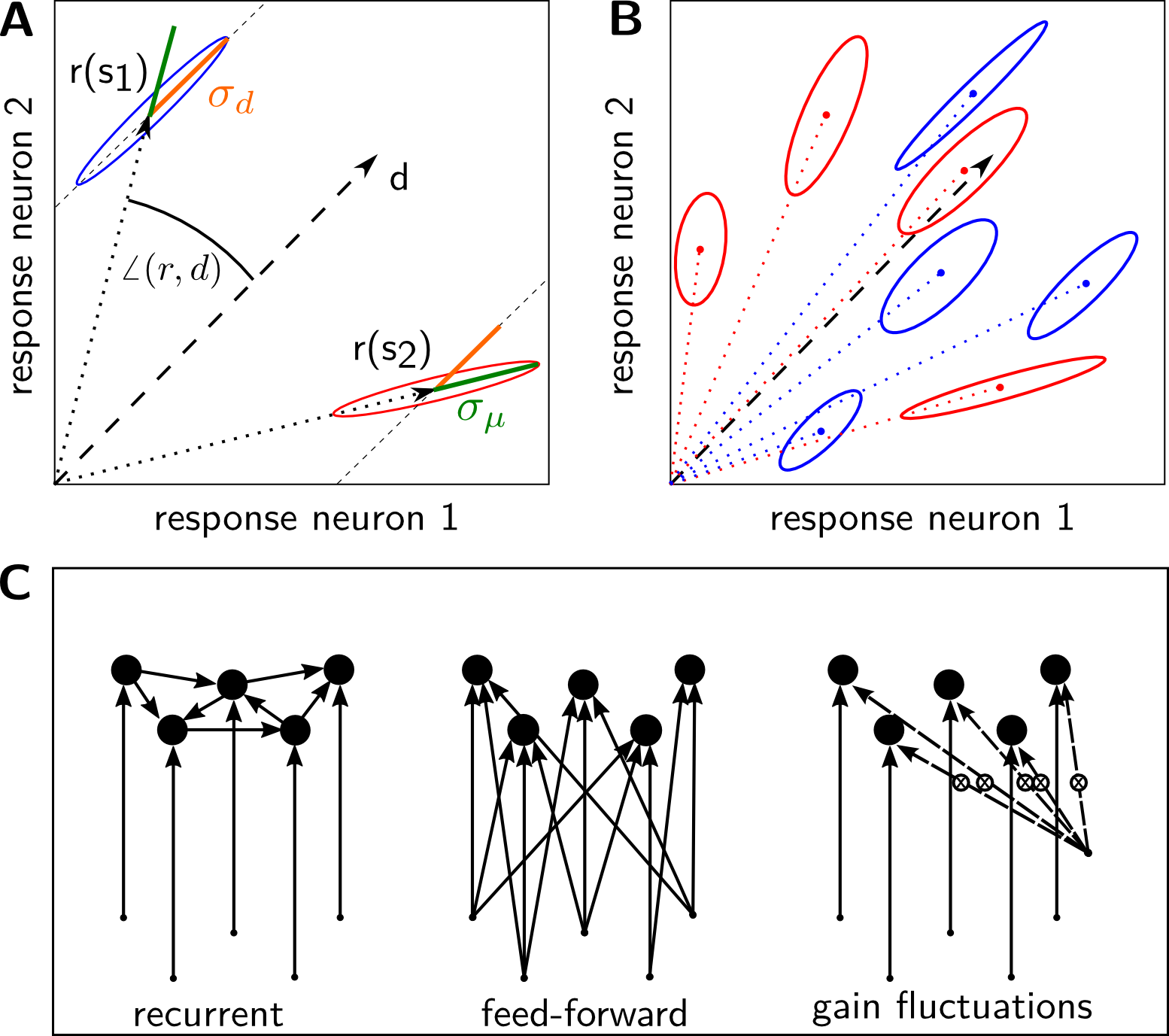}
\end{center}
\caption{ \textbf{Properties of response distributions and network scenarios.} A: Examples for distributions where the variability is to a large
degree in the diagonal direction (response distribution indicated by blue
ellipse), resulting in a large value of $\protect\sigma_d$, and where the
variability is mainly in the direction of the average response (red
ellipse), resulting in a larger value of $\protect\sigma_{\protect\mu}$. 
B: Response distributions for two stimulus ensembles. Either $\sigma_{\mu}$ (red set) or $\sigma_d$ (blue set) remains large across stimuli. A dependence as in the red set appears in a model of shared gain fluctuations, while a dependence as in the blue set may arise in densely connected recurrent or feed-forward networks.
C: Different network layouts that induce correlated activity. Connections (arrows) to and between neurons (dots) vary in strength. Dashed arrows indicate multiplicative effects on firing rates.
\newline
}
\label{fig:sigMuSigD}
\end{figure}

If the vector of population activity in trial $T$ of stimulus $s$ is $\tilde{%
r}(s,T)$, the average response across trials for this stimulus is denoted by 
$r(s)=\langle {\tilde{r}(s,T)}\rangle _{T}$. To measure covariability across
trials in a given stimulus condition between neurons $i$ and $j$, we use the
noise covariances, defined as 
\begin{equation}
C_{ij}(s)=\mathrm{cov}({\tilde{r}_{i}(s,T),\tilde{r}_{j}(s,T)})_{T}=\langle {%
\tilde{r}_{i}(s,T)\tilde{r}_{j}(s,T)}\rangle _{T}-\langle {\tilde{r}_{i}(s,T)%
}\rangle _{T}\langle {\tilde{r}_{j}(s,t)}\rangle _{T}.
\end{equation}%
A measure of the strength of pairwise noise correlations across stimuli is the average correlation coefficient,
\begin{equation}
c_{ij}^{N}=\left\langle {\frac{C_{ij}(s)}{\sqrt{C_{ii}(s)C_{jj}(s)}}}%
\right\rangle _{s}.
\end{equation}%
Here, $\langle . \rangle_s$ denotes the average over all stimuli presented.
This quantity is to be contrasted with the signal correlation, 
\begin{equation}
c_{ij}^{S}=\frac{\mathrm{cov}\bigl({r_{i}(s),r_{j}(s)}\bigr)_s}{\sqrt{\mathrm{var}\bigl({r_{i}(s)}\bigr)_s\mathrm{var}\bigl({r_{j}(s)}\bigr)_s}},
\end{equation}%
 which
measures the similarity of average responses across stimuli. 
As a measure of how the
orientation of the high-dimensional distributions changes across stimuli, we
use the variance of the population activity projected along the direction of
the normalized mean response, $\bar{r}(s)=r(s)/|r(s)|$, namely, 
\begin{equation}
\sigma _{\mu }^{2}(s)=\sum_{ij}C_{ij}(s)\bar{r}_{j}(s)\bar{r}_{i}(s).
\end{equation}%
 This quantity can be
compared to the variance projected along the diagonal direction, $\bar{d}%
=(1,\dots ,1)^{T}/\sqrt{N}$, 
\begin{equation}
\sigma _{d}^{2}(s)=\sum_{ij}C_{ij}(s)\bar{d}_{i}\bar{d}_{j},
\end{equation}%
which corresponds to a uniform averaging of the covariances.
To compare these quantities across stimuli, $\sigma _{\mu }^{2}$ and $\sigma_d^2$ are normalized by the sum of the variances, $\sigAll^2(s)=\sum_iC_{ii}(s)$.
If, for a given stimulus, all neurons are equally active on average, then $\bar r = \bar d$ and $\sigma_d^2 = \sigma_{\mu}^2$. As we show, different circuit models predict different stimulus dependencies of $\sigma_{\mu}^2$ and $\sigma_d^2$. These differences are most apparent
for the stimuli for which the average population response
differs strongly from a uniform population response. A deviation from a uniform response can be measured by the angle between mean response and diagonal, or $\cos (d,r)=\bar{r}^{T}\cdot \bar{d}$.
For a graphical
illustration of these measures, see Fig.\ \ref{fig:sigMuSigD}.

\subsection{Recurrent model of Poisson neurons with noisy input}

\label{methods:hawkes} We model an intrinsically noisy process of spike
generation in neurons, in which the effect of presynaptic spikes on neural
activity is captured through linear modulations of an underlying firing
rate. Specifically, the spike train of neuron $i$ is a realization of an inhomogeneous
Poisson process with a time-dependent firing rate, $\tilde{r}_{i}(t)$,
calculated as 
\begin{equation}
\tilde{r}_{i}(t)=\tilde{r}_{\mathrm{ext},i}(t)+\sum_{j}\int_{0}^{\infty }\tilde{g}%
_{ij}(\tau )\tilde{s}_{j}(t-\tau )d\tau ,  \label{HawkesDynamics}
\end{equation}%
where $\tilde{r}_{\mathrm{ext},i}(t)$ is the external input, $\tilde{s}%
_{j}(t)=\sum_{t^{\prime }}\delta (t-t^{\prime })$ is the recurrent input
from presynaptic spike trains, and the causal coupling kernel, $\tilde{g}%
_{ij}(\tau )$, defines the interactions between neurons in the network. The
external input is an analog quantity which can be viewed as resulting from a
convolution between a linear filter and an incoming spike train from
``non-local'' presynaptic neurons. We treat it as a noisy, but stationary,
signal, i.e., its mean and higher moments do not depend on time. While this
model of interacting point processes \cite{Hawkes1971} defines the full
temporal dynamics of the system, we are only interested in time-averaged rates or, equivalently, spike counts. Their expected values and (co-)variances across realizations can be obtained if the coupling matrix is known.

Assuming stationarity of input spiking and firing rates, we can solve
for the average rate vector, $r=\left( r_{1},\ldots r_{N}\right) $, across
trial realizations, as 
\begin{equation}
r=(\mathbb{I}-G)^{-1}r_{\mathrm{ext}}.  \label{HawkesRates}
\end{equation}%
Here, $\mathbb{I}$ denotes the identity matrix, $G$ is the steady-state
coupling matrix with elements 
\begin{equation}
G_{ij}=\int_{0}^{\infty }\tilde{g}_{ij}(\tau )d\tau ,
\end{equation}%
and $r_{\mathrm{ext}}$ is the average input vector. To describe the correlation in the
activity, we consider the spike count in a long time bin, $\Delta $,
defined as 
\begin{equation}
n_{i}(\Delta )=\int_{t}^{t+\Delta }\tilde{s}_{i}(t)dt.
\end{equation}%
The total input in the time bin is given by%
\begin{equation}
r_{\mathrm{ext}}(\Delta )=\int_{t}^{t+\Delta }\tilde{r}_{E}(t)dt,
\end{equation}%
and is a random variable with a normalized variance, $V_{\mathrm{ext}}\equiv \mathrm{%
var}({r_{\mathrm{ext}}(\Delta )})/\Delta $ (across choices of different time bins).
From spectral properties of the covariance \cite{Hawkes1971} and the
relation between covariance and count correlations \cite%
{Brody1999,Pernice2011}, it can be shown that the count covariance
normalized by the time bin duration, $C$, with elements 
\begin{equation}
C_{ij}\equiv \lim_{\Delta \rightarrow \infty }\frac{\mathrm{cov}\bigl(%
n_{i}(\Delta ),n_{j}(\Delta )\bigr)}{\Delta },
\end{equation}%
is given by the matrix 
\begin{equation}
C=(\mathbb{I}-G)^{-1}\bigl(D[r]+D[V_{\mathrm{ext}}]\bigr)(\mathbb{I}-G^{T})^{-1}.
\label{HawkesCovs}
\end{equation}%
The term $D[r]$ denotes a diagonal matrix with elements $D[r]_{ij}=r_{i}%
\delta _{ij}$, and similarly $D[V_{\mathrm{ext}}]_{ij}=\delta _{ij}V_{\mathrm{ext},i}$. The effect
of the recurrent network interactions is reflected in the transfer matrix, 
\begin{equation}
B\equiv (\mathbb{I}-G)^{-1}.
\end{equation}%
In Eq.\ (\ref{HawkesCovs}), one component of the covariance arises from the
variance of the external input, $V_{\mathrm{ext}}$. Although the intrinsic noise from
the spike-generating Poisson process is dynamical, its effect is that of an
additive contribution to the external variance (see also \cite{Grytskyy2013}%
). The dependence on the rate vector, in Eq.\ (\ref{HawkesCovs}), results
from the Poisson character of the noise, in which the variance equates the
mean.\newline

In the experimental data set, 
 firing rates are measured relative to
spontaneous activity. To take this into account, an offset, $a$, can be
added to the rates in Eq.\ (\ref{HawkesCovs}). Henceforth, we examine the
model through Eqs.\ (\ref{HawkesRates}) and (\ref{HawkesCovs}) for the mean
neural activities and their covariances, respectively.

\subsubsection{Random effective network and stimulus ensemble}
\label{sec:effRandom}

\label{randEffModel} The rates and covariances given in Eqs.\ (\ref%
{HawkesRates}) and (\ref{HawkesCovs}) depend on the recurrent coupling
matrix within the network, $G$, only via effective couplings defined by the
transfer matrix, $B=(\mathbb{I}-G)^{-1}$.
In numerical calculations, we choose the entries of $G$ independently from a
normal distribution with positive mean; this affords a specific
structure to the matrix $B$. The empirical mean and variance of the elements of this matrix will be called $\langle {B}\rangle $ and $\mathrm{var}({B})$. In a number of analytical calculations, for the sake of
simplicity, we neglect the specific structure of $B$ and model it as a random matrix with elements independently drawn from a normal distribution with corresponding mean and variance. The elements of $B$ are not direct, but effective connections, because they reflect also the effects of indirect connections in the network, hence the expression ``random effective network.''
The variability of effective
connections can be quantified by the ratio $\rho =\mathrm{var}({B}%
)/\langle {B}\rangle ^{2}$. 

The stimulus ensemble is modeled as a set of random vectors, $r_{\mathrm{ext}}(s)$. We assume that the
elements of these vectors are  independent across neurons and stimuli, and normal, and we call the mean and variance of the corresponding normal distribution $\langle {%
r_{\mathrm{ext}}}\rangle $ and  $\mathrm{var}({r_{\mathrm{ext}}})$. 
 The relative variability of the inputs is defined as  $%
\rho _{E}=\mathrm{var}({r_{\mathrm{ext}}})/\langle {r_{\mathrm{ext}}}\rangle$. This variability is high if both excitatory and inhibitory inputs are present, and,
together with the network parameters, it  
determines the variability of the average responses across stimuli. 
In numerical calculations, we set $V_{\mathrm{ext},i}(s)=|r_{\mathrm{ext},i}(s)|$, as dictated by
a Poisson process. We allow for input \textit{signal} correlations, i.e.,
the average inputs to a pair of neurons can come with a non-vanishing
correlation coefficient, $c_{\mathrm{in}}=\left\langle {%
r_{\mathrm{ext},i}(s)_{i},r_{\mathrm{ext},j}(s)}\right\rangle _{s}$.

\subsection{Models of correlated activity from shared inputs or gain
modulations}

\label{methods:alternative} Correlations in the activity of neurons can have
other origins besides recurrent connections. In parallel with the recurrent
network model, we consider two alternative prototypical models in which
correlations originate from shared inputs or from shared gain fluctuations,
respectively. The three different scenarios are illustrated in Fig.\ \ref{fig:sigMuSigD}C. 
 In Appendix \ref{app:commonFramework}, we show how, formally,
these models can also be cast as special cases of the recurrent model.

\subsubsection{Feed-forward model with shared input}
We consider a simple, two-layer, feed-forward network, in which $N$ output neurons
receive inputs from $N$ independent presynaptic neurons with spike trains $%
\tilde{s}_{j}(t)$. 
The contributions of the input neurons to the firing rates of the output neurons are determined by  feed-forward connection kernels $\tilde f_{ij}(\tau)$, so that the observed firing rates are 

\begin{equation}
\tilde{r}_{i}(t)=\sum_{j}\int_{0}^{\infty }\tilde{f}%
_{ij}(\tau )\tilde{s}_{j}(t-\tau )d\tau.
\end{equation}%
As we show in Appendix \ref{app:commonFramework}, this scenario represents a special case of the framework described in Section \ref{methods:hawkes}, so that firing rates and count covariances can be calculated correspondingly. We define the feed-forward coupling matrix $F$ by its elements, $F_{ij}=\int_{0}^{\infty }\tilde{f}_{ij}(\tau )d\tau$. With the time-averaged firing rates of the external input for stimulus $s$, $r_{\mathrm{ext}}(s)$,
 the average spike
counts of the output neurons are given by
\begin{equation}
r(s)=Fr_{\mathrm{ext}}(s).  \label{FFNrates}
\end{equation}%
If the spike trains of the input neurons are Poisson processes, the input variance is equal to the rate,  $V_{\mathrm{ext}}=r_{\mathrm{ext}}$, but we will allow for more general inputs. The count covariances are then given by
\begin{equation}
C=FD[V_{\mathrm{ext}}]F^{T}+D[r].  \label{covsFFN}
\end{equation}%
We denote again by $D[r], D[V_{\mathrm{ext}}]$ diagonal matrices with diagonal elements given by the vectors $r, V_{\mathrm{ext}}$.
The first term describes covariances resulting from the combined shared
inputs. The second term results from the contribution to the count variances of the Poisson spike generation in output neurons, which is independent across neurons.

If rates are observed only up to a stimulus-dependent offset, $a$, 
 Eq.\ (\ref{covsFFN}) has to be replaced by 
\begin{equation}
C=FD[V_{\mathrm{ext}}]F^T+D[r+a].
\label{eq:covsFFNoffset}
\end{equation}
This scenario can be compared directly to the random effective recurrent
network model, if one uses $F=B$ for an identical ensemble of stimuli (see
Section \ref{randEffModel}). 
Comparing Eq.\ (\ref{eq:covsFFNoffset}) to Eq.\ (\ref{HawkesCovs}), we note that the
difference between the two models is that the internally generated noise 
 in the feed-forward model, $D[r+a]$,  is uncorrelated and contributes only to
the variances, while in the recurrent network it is filtered by the network
and therefore correlated. 

\subsubsection{Model with shared gain fluctuations}

In addition to recurrence and shared inputs, shared gain fluctuations have also been proposed as a source of correlation in  neural populations by various authors  \cite{Goris2014, Okun2015, Franke2015, Ecker2015}. 
 In this model, the firing rate of the neural population is given by the product of a constant, stimulus dependent vector, $r(s)$, and a scalar fluctuating signal, $\epsilon (t)$. We set the time-averaged value of the fluctuations to unity, $\langle \epsilon(t) \rangle_t=1$, so that the time-averaged firing rate is $r$.
 The resulting count
covariance matrix is given by
\begin{equation}
C=D[r]+rr^{T}V_{\mathrm{ext}}  \label{FF1covs}, 
\end{equation}%
where the scalar variance, $V_{\mathrm{ext}}$, reflects the strength of the variation of the fluctuating signal.
In other words, the firing rate of a Poisson neuron is modulated multiplicatively by a fluctuating ``gain'' signal. One consequence is that the covariance of a neuron pair is proportional to the product of their firing rates.
The general model described in Section \ref{methods:hawkes}  reduces to this model if for any given stimulus  the activity of the population results from a single input neuron projecting to the output neurons with corresponding weights (see Appendix \ref{app:commonFramework}).

Here, the average responses  can be chosen freely. When comparing
this scenario to the recurrent and feed-forward scenarios, we use the ensemble
of average responses resulting from the corresponding network scenario. If
the stimulus dependence of the firing rates is measured relative to an
offset $a$, the covariances are given by 
\begin{equation}
C=D[r+a]+(r+a)(r+a)^{T}V_{\mathrm{ext}}.  \label{sharedCovsOffset}
\end{equation}%

\subsection{Stimulus discriminability}

\label{sec:discrimination} In order to evaluate the influence of correlation
on neural coding, we examine the discriminability of a set of stimuli from
the population activity. In the models, a stimulus, $s$, 
 evokes activity characterized by an average response
vector, $r(s)$, and a covariance matrix, $C(s)$. 
We seek a simple measure for evaluating the possibility of attributing a given population response to one of two discrete stimuli, $s_1$ and $s_2$, unambiguously. For this, we assume that the high-dimensional distributions of responses are Gaussian, and project the two distributions on a single dimension where we calculate the signal-to-noise ratio as our measure.

The best projection follows from Fisher's linear discriminant analysis: the most informative linear
combination of neuron responses, denoted by $w\cdot r$, is achieved if the
vector $w$ points in the most discriminant direction, 
\begin{equation}
\label{eq:mostDD}
w=\Bigl(C(s_{1})+C(s_{2})\Bigr)^{-1}\Bigl(r(s_{1})-r(s_{2})\Bigr).
\end{equation}
The mean and variance of the projected distributions onto the normalized
direction, $\bar{w}$, are $\bar{w}^{T}r(s_{i})$ and $\sigma _{s_{i}}^{2}=\bar{w%
}^{T}C(s_{i})\bar{w}$, for $i\in \{1,2\}$. 
We can then define the \textquotedblleft signal-to-noise
ratio,\textquotedblright\ as 
\begin{equation}
\label{eq:sigToNoise}
S=\frac{\big|\bar{w}^{T}\cdot \bigl(r(s_{1})-r(s_{2})\bigr)\big|}{\sigma _{s_1}+\sigma
_{s_2}}.  
\end{equation}%
 Larger values of $S$ correspond to better discriminability. 

To quantify the effect of correlation on discriminability, we compare the
quantity $S$, defined in Eq.\ (\ref{eq:sigToNoise}), with the quantity $S_{\text{shuffled}%
}$ obtained from a ``shuffled data set'', 
 in which
responses are shuffled across trials (in experimental data) or off-diagonal
covariance elements, $C_{i\neq j}(s)$, are set to 0 (in models). A ratio $S_{%
\text{shuffled}}/S\ $smaller than unity indicates that correlation is
beneficial to discriminability.

We note that $S$ is an approximation for a measure based on the optimal linear classifier, for which the threshold separating the two one-dimensional projected distributions has to be calculated numerically. A measure akin to $S$ used in similar contexts is the linear Fisher information \cite{Abbott1999}, valid for continuous stimuli. 
An advantageous property of $S$ is its invariance under linear transformations: if all responses, $r$, are fed into another network whose output, $Br$, results from a product with an invertible matrix $B$, $S$ does not change. This obtains because the mean responses are transformed into $Br(s)$,  while the covariances are transformed into $BC(s)B^T$. Intuitively, a simple matrix multiplication is accommodated for by a corresponding change in the most discriminative direction $w$ on which we project.

\subsection{Two-population model}
\label{sec:twopop}
In order to develop an intuitive understanding of the way in which
correlations generated in a recurrent network influence coding, we examine a
highly simplified model. The recurrent network consists of two excitatory
sub-populations, labeled $E$ and $E^{\prime }$, each made up of $N$ neurons.
We ask to what extent their activity discriminates two stimuli when these
elicit preferential responses in the two sub-populations, respectively.

The activity in the network is determined by Eqs.\ (\ref{HawkesRates}) and (%
\ref{HawkesCovs}). For the sake of simplicity, we can set the external variance
to zero, such that the input to the network is fully defined by its mean
input, $r_{\mathrm{ext}}$. We assume some further simplifications, for the sake of
calculational ease:\ each neuron in sub-population $L$ projects to a fixed
number, $n_{KL}$, of neurons in sub-population $K$; 
all non-zero coupling weights are identical, $%
G_{ij}=g_{E}$.
Additionally, all neurons in the same population receive identical external input. Then, the two-component vectors $R_{\mathrm{ext}}(s_{1})=(1+\Delta ,1)^{T}$ and $R_{\mathrm{ext}}(s_{2})=(1,1+\Delta
)^{T}$ define a pair of stimuli. Each component denotes the  input to each neuron in the corresponding population.

We reduce the dimensionality of the system by considering the population activity on a macroscopic level.
 The average response of population $K$ across
trials is defined as 
\begin{equation}
R_{K}=\sum_{k\in K}r_{k},
\label{popRates}
\end{equation}
and its trial-to-trial variability is described by the population covariance
matrix, with elements defined as
\begin{equation}
\Sigma _{KL}=\sum_{k\in K,l\in L}C_{kl}.
\label{popCovariances}
\end{equation}
 It turns out that these macroscopic quantities depend only on the overall number of connections between populations, not on a specific network realization. For example, the sum of the rates of neurons in one population depends only on the sum of the inputs to all neurons, but not on how these inputs are distributed among the neurons. 
This is a consequence of the linearity of the dynamics and the assumption that each neuron has a fixed number of output connections, as we show in detail in Appendix \ref{app:popModel}. This idea was already applied in Refs.\ \cite{Pernice2011, Trousdale2012}.

More precisely, we define the coupling within each population, $\Gamma
_{s}=n_{EE}g_{E}=n_{E^{\prime }E^{\prime }}g_{E}$,  and across populations, $\Gamma _{c}=n_{EE^{\prime }}g_{E}=n_{E^{\prime }E}g_{E}$, which make up a population coupling matrix
\begin{equation}
\Gamma =%
\begin{pmatrix}
\Gamma _{s} & \Gamma _{c} \\ 
\Gamma _{c} & \Gamma _{s} \\ 
\end{pmatrix}.
\label{popCoupling}
\end{equation}%
Applying the definition in Eqs.\ (\ref{popRates}) and (\ref{popCovariances}), we can rewrite the microscopic Eqs.\ (\ref{HawkesRates}) and (\ref{HawkesCovs}) as population equations, 
\begin{equation}
R(s)=(\mathbb{I}-\Gamma )^{-1}NR_{\mathrm{ext}}(s)
\label{eq:popRates}
\end{equation}%
and 
\begin{equation}
\Sigma (s)=(\mathbb{I}-\Gamma )^{-1}D[R(s)](\mathbb{I}-\Gamma )^{-1}.
\label{eq:popCovs}
\end{equation}%
The population transfer matrix is defined by $P=(\mathbb{I}-\Gamma )^{-1}$.
 Up to a factor $N$, these equations are equivalent to the microscopic equations in the restricted context of the two-population model. 

\section{Results}

We discuss the characteristics of the noise statistics that emerge from
neural dynamics in models of recurrent networks, feed-forward networks, and
networks with global gain fluctuations. We also look for coarse statistical
signatures of each of these three model structures, which can then be
compared with data. In this perspective, we analyze cortical recordings
in  mouse. These recordings were made in the auditory cortex, while the animal
was stimulated with musical tones \cite{Bathellier2012}.
We then discuss the
implications of the neural noise statistics for stimulus encoding. Finally,
we present a highly simplified model of network dynamics, paired down to
include only a few parameters, in order to develop an intuition of the link
between network structure and stimulus encoding.

\subsection{Population signatures of noise statistics in a recurrent network
model}

\label{sec:varRecNet}
\begin{figure}[H]
\begin{center}
\includegraphics[width=13.2cm]{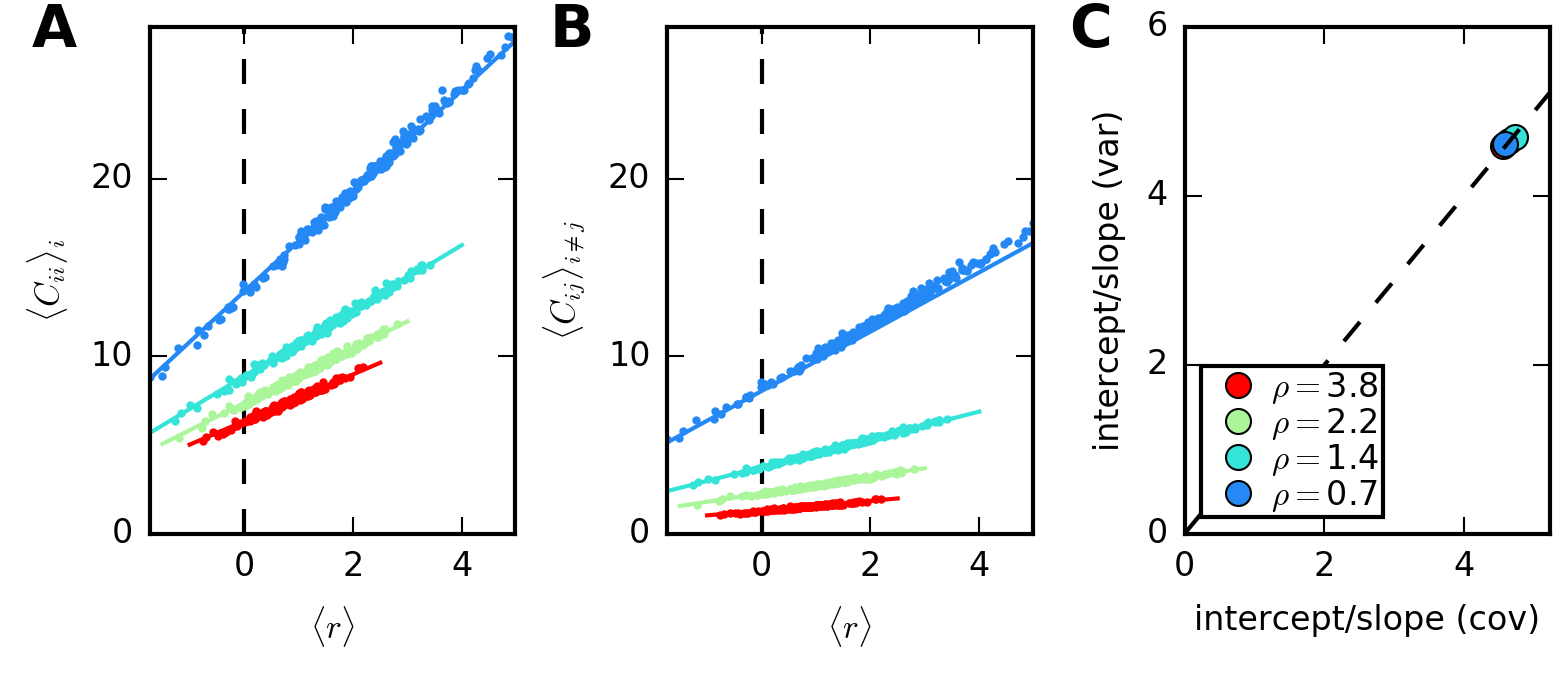}
\end{center}
\caption{ \textbf{Variability in the recurrent network model.} A: Relation
between response and variance averaged across population for randomly
chosen stimulus vectors (each dot one stimulus). Four networks are generated
with normally distributed connection strengths. Larger $\protect\rho$,
calculated from effective connections in the transfer matrix $B$, indicate
larger variability of effective weights. Lines indicate analytic
predictions, Eqs.\ (\protect\ref{eq:varScalingRec}) and (\protect\ref%
{eq:covScalingRec}). B: Same for covariances, averaged across all neuron
pairs. C: The ratio of slope and intercept of linear fits corresponds to the
strength of the input variance, which is identical for all networks, and is
consistent for variances and covariances. See Appendix \protect\ref%
{app:figDetails} for further details and the numerical parameters. \newline
}
\label{fig:randomModelSigs}
\end{figure}

 One population feature which distinguishes the
possible mechanisms of generation of noise correlation is the relation
between the population-averaged response, $\langle {r}\rangle (s)=\langle {%
r_{i}(s)}\rangle _{i}=\frac{1}{N}\sum_{i}r_{i}(s)$, on the one hand, and the
population-averaged variance, $\langle {C_{ii}}\rangle _{i}(s)=\frac{1}{N}%
\sum_{i}C_{ii}(s)$, or the noise covariance averaged across pairs $\langle {%
C_{ij}(s)}\rangle _{i\neq j}=\frac{1}{N(N-1)}\sum_{i\neq j}C_{ij}$, 
 on the other hand.
In this section, we analyze these relations for Poisson neurons in a recurrent network with random effective connections, as described in Section \ref{sec:effRandom}.
Using Eq.\ (\ref{HawkesCovs}), which reads $C_{ij} = \sum_{k}B_{ik}B_{jk}(r_k(s)+a+V_{\mathrm{ext},k})$ for the pairwise covariances, we derive the expression
\begin{equation}
\langle {C_{ii}}\rangle _{i}(s)\approx N\langle {B^{2}}\rangle \Bigl(\langle 
{r}\rangle (s)+a+\langle {V_{\mathrm{ext}}}\rangle\Bigr) 
\label{eq:varScalingRec}
\end{equation}%
for the average variance, and the expression
\begin{equation}
\langle {C_{ij}}\rangle _{i\neq j}(s)\approx N\langle {B}\rangle ^{2}\Bigl(%
\langle {r}\rangle (s)+ a + \langle {V_{\mathrm{ext}}}\rangle\Bigr)
\label{eq:covScalingRec}
\end{equation}%
for the average covariance. (See Appendix \ref{app:recurrent} for mathematical details and Fig.\ \ref{fig:randomModelSigs} for numerical results.) The quantity $\langle {B}\rangle =\frac{1}{N^2} \sum_{ij}B_{ij}$ denotes the average strength of the effective connections in the network across neuron pairs and,  correspondingly,  $\langle {B^{2}}\rangle = \frac{1}{N^2}\sum_{ij}B^2_{ij}$ the average of their square. The variance of the input, averaged across neurons, is $\langle {V_{\mathrm{ext}}}\rangle$, and $a$ denotes a potential constant offset in the observed firing rates.

 The variance of a Poisson variable is proportional to its mean. This contribution from the Poisson spike generation is reflected by the term $\langle {r}\rangle $ in the two equations and is the reason for the linear relation between average response and average variance/covariance - for non-Poisson spike generation,
these relations would be non-linear. As we see, not only the external noise, $\langle {V_{\mathrm{ext}}}\rangle$, but also the internally generated noise  are amplified by the recurrent connections.
The amplification factor, or the slope of the linear relation, $N\langle {B}\rangle^2 $ in Eq.\ (\ref{eq:covScalingRec}), is smaller for the covariances, where it depends only on the mean strength $\langle {B}\rangle$ of the effective connections. For the variances, expressed in Eq.\ (\ref{eq:varScalingRec}), the averaging in the slope $N\langle {B^{2}}\rangle $ is carried out on the square of the effective connections. The slope is therefore larger, and depends both on the average strength of connections and their variability.
In the relations for both average
population variance and covariance, the ratio between the intercept and the
slope is $\langle {V_{\mathrm{ext}}}\rangle +a$. Up to a possible offset, $a$,  that can result from a measurement of firing rates relative to a constant baseline activity, it represents the strength of the external noise.

In summary, in the recurrent network model, both internally generated noise and 
external noise are amplified (or attenuated) by the recurrent connections, and the parameters
in the relationship between (co-)variances and population response can be related
to  network and input parameters.

\subsection{Population signatures of noise statistics in feed-forward network
models}

\label{sec:theoryAltern} In a feed-forward network, both shared input units
and global gain fluctuations can yield noise correlations \cite%
{Goris2014,Ecker2015, Franke2015}. 
Here, we compare
the population signatures of the noise statistics in the case of these two
alternatives with those from a recurrent model. This serves to
point out differences between the outputs of the various models as well as
to motivate the data analysis to which we turn below.

\begin{figure}[H]
\begin{center}
\includegraphics[width=13.2cm]{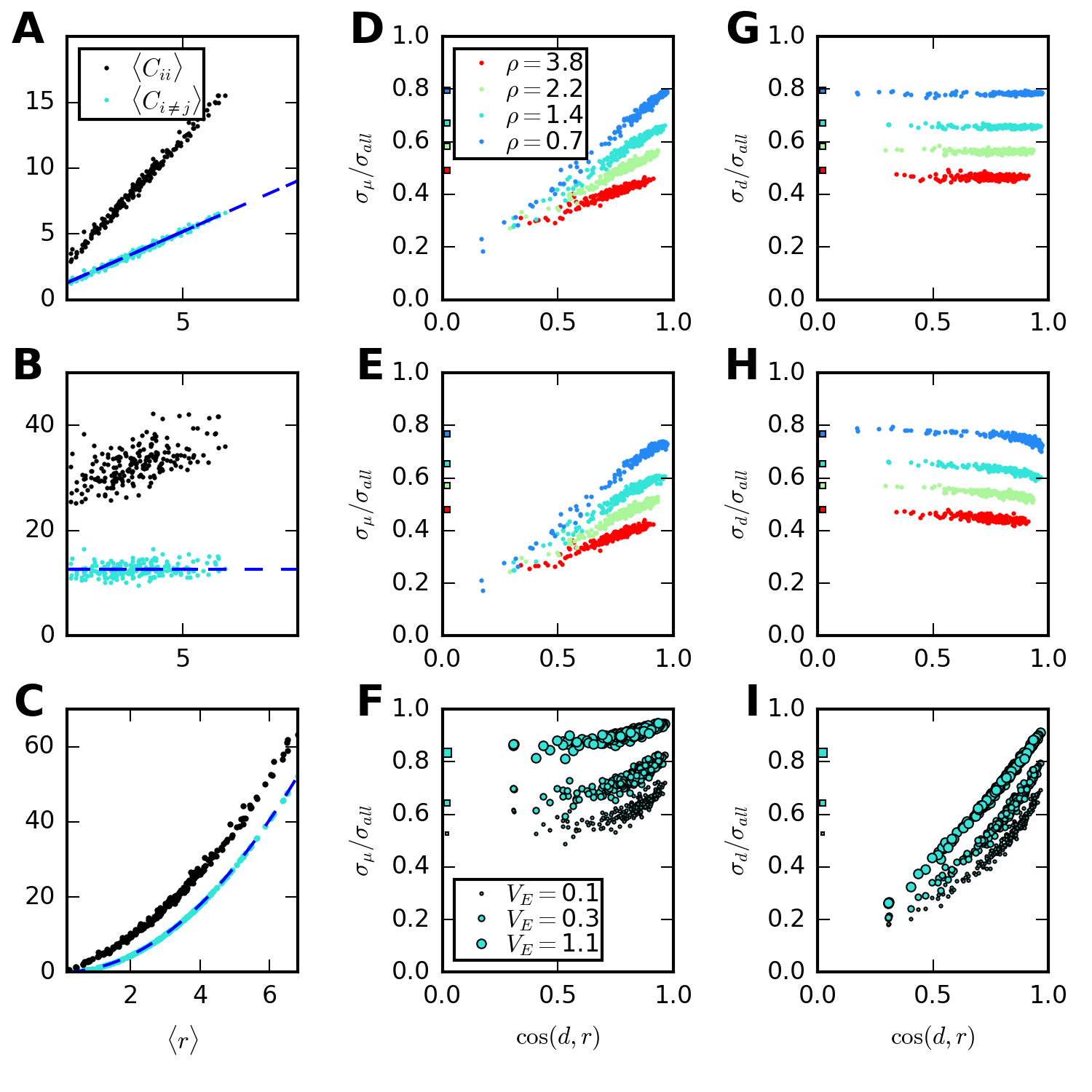}
\end{center}
\caption{ \textbf{Response-covariance relations depend on origin of
correlations. } A-C: Average population response $\langle {r} \rangle$ versus average
(co-)variances in recurrent network (top), feed-forward network (middle) and
gain-fluctuation model (bottom). Each dot corresponds to a random stimulus, blue
dashed lines indicate analytic results, Eqs.\ (\protect\ref{eq:covScalingRec}%
)-(\protect\ref{eq:covScalingGain}). D-F: Dependence of normalized
variability projected on mean direction on response direction for the three
scenarios. In D,E colors indicate results for different networks, in F size
of markers indicates strength of gain fluctuations. Square markers on the
left indicate numerical value of $\protect\sqrt{c_N}$. G-I: Same for the
normalized variability projected on diagonal direction. See Appendix for
further details and the numerical parameters. }
\label{fig:covMeanDep}
\end{figure}

The three scenarios differ in the scaling of the population-averaged
covariance, $\langle {C_{ij}}\rangle _{i\neq j}(s)$, with respect to the
population response, $\langle {r}\rangle (s)$ (Fig.\ \ref{fig:covMeanDep}).
We already demonstrated in the previous section that both variances and
covariances scale linearly with the population response in the recurrent network. 
In a feed-forward network, pairwise covariances are given by Eq.\ (\ref{covsFFN}), which can be rewritten as
\begin{equation}
 C_{ij}=\delta_{ij}(r_{i}+a)+\sum_kF_{ik}F_{jk}V_{\mathrm{ext},k}.
\end{equation}
 Note that, in contrast to the recurrent network, the neural firing rates, $r_i$, only affect diagonal entries, and therefore the variances.
The
average covariance can be expressed as
\begin{equation}
\langle {C_{ij}}\rangle _{i\neq j}(s)\approx N\langle {F}\rangle ^{2}\langle 
{V_{\mathrm{ext}}}\rangle (s),  \label{eq:covScalingFF}
\end{equation}%
 and does not directly depend on the average population response, $\langle r \rangle(s)$ (see Appendix \ref{app:FFmodel}). If there are excitatory as well as inhibitory inputs, also the average input variance, $\langle V_{\mathrm{ext}} \rangle$, can be uncorrelated to the population response:  for Poisson external input, in each input channel $i$ the variance is as large as the mean input, $V_{\mathrm{ext},i} = |r_{\mathrm{ext},i}|$. Nonetheless, because the variance is always positive, including for negative input, the average variance $\langle V_{\mathrm{ext}} \rangle$ can be decorrelated from the average input, $\langle r_{\mathrm{ext}} \rangle$, and thus from the average output, $\langle r \rangle$. Intuitively, because inhibitory channels add variance, but decrease the average input, the combined input does not have to be proportional to the combined variance. The extent to which this is true depends on the balance between positive and negative elements in $r_{\mathrm{ext}}$, which we measure by the input variability $\rho_E = \mathrm{var}({r_{\mathrm{ext}}})/\langle {r_{\mathrm{ext}}}\rangle$.

In a model network with global gain fluctuations, from Eq.\ (\ref{FF1covs}), the covariances are given by 
\begin{equation}
C_{ij}=\delta_{ij}r_{i}+r_{i}r_{j}V_{\mathrm{ext}}.
\end{equation}
 Again averaging only across the off-diagonal elements, one finds that the average covariances scale
quadratically with the population response, as
\begin{equation}
\langle {C_{ij}}\rangle _{i\neq j}=V_{\mathrm{ext}}\langle {r}\rangle ^{2}.
\label{eq:covScalingGain}
\end{equation}%
This is a direct consequence of the fact that pairwise covariances are proportional to the product of the firing rates of the pair of neurons, which holds for large $N$ independent of the distribution of average responses. By contrast, as is apparent from the expression for $C_{ij}$, the individual variances and the population averaged variance each have both a quadratic and a linear contribution in $r$ (and $\langle r \rangle$, respectively) (see also e.g.\ \cite{Goris2014}).
 (Further details are provided in Appendix \ref{app:ThreeModels}.)

Until now, we examined how the average covariance $\langle C_{ij}\rangle_{i\neq j}$ changes across stimuli. Further differences between the different scenarios can be related to the detailed structure of $C$, which determines the shape of the response distributions.
By shape, we refer to the geometric orientation and extent of the multi-dimensional ellipsoid cloud that corresponds to the response distribution in the space spanned by the responses of the individual neurons.
A two-dimensional sketch of such an ellipse, and the geometric interpretation of the quantities we use is depicted in Fig.\ \ref{fig:sigMuSigD}A (see also Methods \ref{sec:methodsRepVar}).
In words, the way the orientation of the response distribution
changes with the average response varies between the three scenarios: the
long axis of the  ellipsoid indicating the direction of
the largest variance is aligned along the diagonal for the feed-forward and the recurrent network models,
but in the direction of the mean response for the gain fluctuation model
(Fig.\ \ref{fig:sigMuSigD}B). To capture this relation quantitatively, we
consider the projected variances, $\sigma^2_{\mu}$ and $\sigma^2_d$ (see Fig.\ \ref{fig:covMeanDep} and methods for
details). There, the \textquotedblleft diagonal direction\textquotedblright\ 
corresponds to a response where all neurons are equally active.

Quantitatively, we find that in the recurrent and feed-forward network models with
relatively strong correlations, there is always a large fraction of variance
in the diagonal direction, independent of the stimulus.  The variance projected along the direction of the average response is only large if the latter happens to be similar to the diagonal direction. By contrast, in the gain fluctuation model, the variance projected along the direction of the average response is approximately constant across stimuli. This behavior of the gain fluctuation model is due to the fact that the variability across trials is multiplicative on the mean response, and hence variability is large in this direction for all stimuli.

In the network models, strong correlations result from effective interactions between neurons or from shared input. Each neuron in the recurrent network, or each input channel in the feed-forward network, is effectively connected to a large part of the remaining neurons and contributes to their correlations. Because our networks are not balanced, these contributions are mostly positive, even for inhibitory input channels. Thus, across trials, responses of all neurons are strongly correlated and the trial-to-trial fluctuations tend to be of similar size for each neuron for a given stimulus. By contrast, the variability for any given neuron varies more strongly across stimuli.

\subsection{Relation between signal correlation and noise correlation in
recurrent network models}

In the simple scenario we analyze here, average responses and covariances
are related because they both depend on the network architecture. By
expressing signal correlations, $c_{ij}^{S}$, and the noise correlations, $%
c_{ij}^{N}$, in terms of network and input parameters, we can derive a
relation between these two sets of quantities (Fig.\ \ref%
{fig:randomModelCorrs}). For a random transfer matrix, $B$, all neuron pairs
are statistically equivalent, and the strength of correlations can be
characterized by the average of the pairwise signal and noise correlations.
In our recurrent network model,
the network
architecture affects noise and signal correlations through an effective
quantity, namely the signal-to-noise ratio of the elements of the
transfer matrix, $\rho = \mathrm{var}({B})/\langle {B}\rangle ^{2}$: 
\begin{equation}
c_{N}\equiv \langle {c_{ij}^{N}}\rangle _{i\neq j}\approx \frac{1}{1+\rho },  \label{eq:cnoise}
\end{equation}
\begin{equation}
c_{S}\equiv \langle {c_{ij}^{S}}\rangle _{i\neq j}\approx \frac{1+(N-1)c_{\mathrm{in}}%
}{1+\rho +(N-1)c_{\mathrm{in}}}  \label{eq:csig}
\end{equation}%
(see Appendix \ref{app:recurrent}, Eqs.\ (\ref{Eq:noiseCorr}) and (\ref{Eq:signalCorr})). Here, $c_{\mathrm{in}}$ denotes the strength of signal correlations in the input across stimuli.
 Both signal and noise correlations are larger in
networks with more homogeneous entries (smaller $\rho $).  We already showed in Section \ref{sec:varRecNet} that noise covariances, $\langle C_{ij}\rangle_{i\neq j}$, depend on the average strength of the effective connections, while the variances, $\langle C_{ii}\rangle_i$, depend on $\var{B}$  as well.
 If the average
input for pairs of neurons across stimuli is uncorrelated (no input signal
correlations, $c_{\mathrm{in}}=0$), noise and signal correlations are identical on
average. However, due to a prefactor which grows with system size ($N-1$ in
Eq.\ (\ref{eq:csig})), even weak input signal correlations are strongly amplified and
can yield a large effect. Because noise in the population response to a
given stimulus is produced internally in the process of spike generation,
noise covariances are unaffected by this mechanism, so that strong signal
correlations can coexist with weaker noise correlations.

These relations depend on network parameters and hold for the average correlations (across pairs in a network). The pairwise noise and signal correlation coefficients, $c_{ij}^N$ and $c_{ij}^2$,  can vary widely, but numerical calculations indicate that they are correlated in a given network (across pairs) as well.

\begin{figure}[H]
\begin{center}
\includegraphics[width=13.2cm]{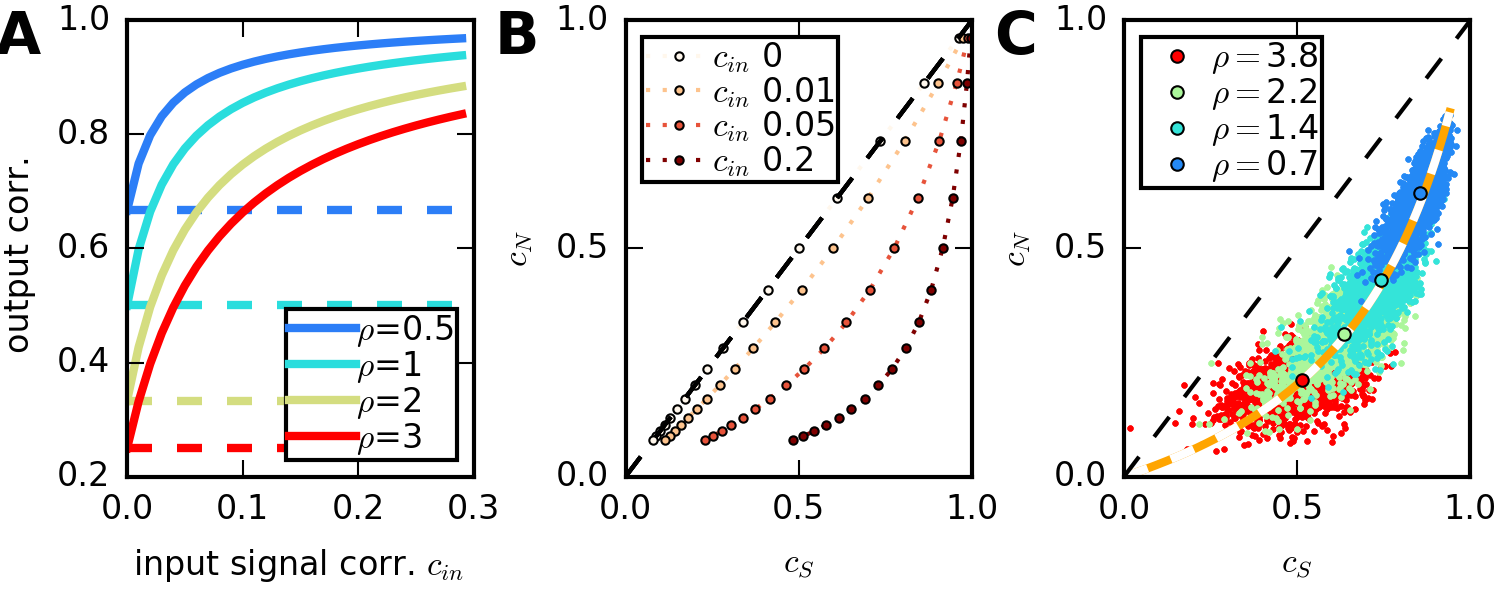}
\end{center}
\caption{ \textbf{Relation between signal and noise correlations in the recurrent
network model. } A: Dependence of average signal correlations, $c_S$, (continuous lines)
 and average noise correlations, $c_N$, (dashed lines) on
input correlation and and network properties, Eqs.\ (\protect\ref{eq:cnoise}%
), (\protect\ref{eq:csig}). Low variability of the transfer matrix elements, $\rho$,
increases $c_N$ and $c_S$. Input signal correlations
affect only $c_S$. B: Same data, average signal versus noise
correlation. C: Scatter plot of all pairwise signal versus pairwise noise
correlations (dots) in five network realizations  for $c_{\mathrm{in}}=0.05$. Circles
indicate network average across pairs, orange dotted line corresponds to
analytic expressions shown in B for $c_N$ and $c_S$. }
\label{fig:randomModelCorrs}
\end{figure}

\subsection{Stimulus discrimination in a recurrent network model}

\begin{figure}[H]
\begin{center}
\includegraphics[width=11.25cm]{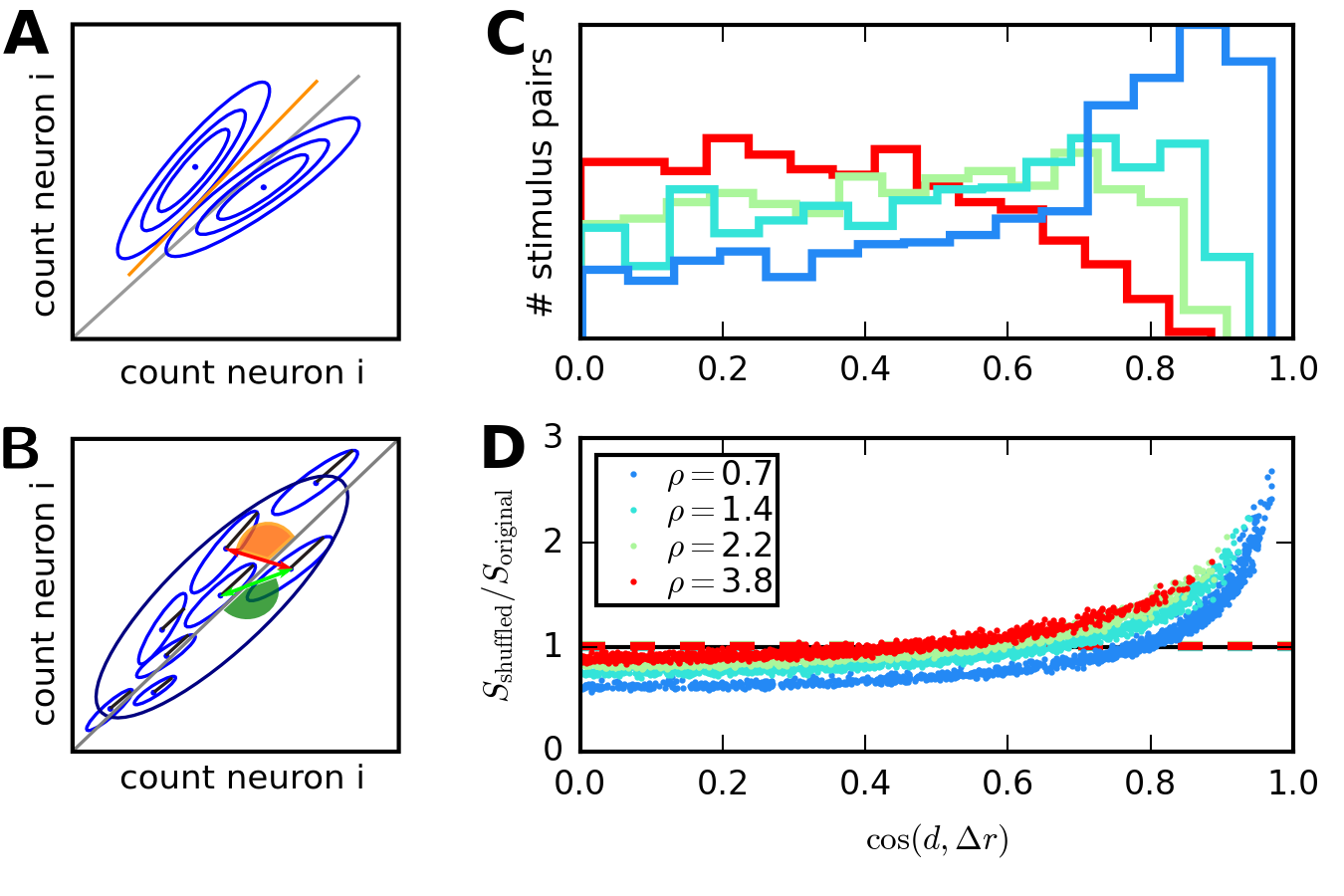}
\end{center}
\caption{\textbf{Covariances affect stimulus discriminability} A: Sketch of
response distributions for two neurons and two stimuli (blue ellipses).
Neural responses are correlated for both stimuli, so that main axis of
variability is parallel to linear separator (orange line). B: Larger
ensemble of stimuli for correlated neurons (large variance
along diagonal directions). The
effect of correlations is favorable for stimulus pairs where difference in
means is orthogonal to diagonal (red arrow), or unfavorable
when difference in means tends to be parallel to diagonal (green arrow). C:
Distribution of cosines of angle between diagonal and difference in means
across stimulus pairs in recurrent networks with different parameter $%
\protect\rho$. In networks with low $\protect\rho$, unfavorable angles are
more frequent. D: Ratio between discriminability for correlated and
shuffled distributions, dots correspond to stimulus pairs. Lower values
indicate that correlations are more beneficial. Dashed horizontal lines indicate
average across all stimulus pairs. }
\label{fig:randomModelStimDisc}
\end{figure}

Since the network architecture affects both signal and noise correlation, it
is natural to ask how the discriminability of stimuli is affected in turn,
as it depends on both quantities. Indeed, noise correlations can affect the
coding properties of a population appreciably \cite{Averbeck2006a}.
 Noise
correlations are referred to as favorable for the discrimination of a pair
of stimuli when the shapes of the two correlated response distributions
(corresponding to the two stimuli) is such that variability occurs
predominantly in a direction orthogonal to the informative direction (Fig.\ %
\ref{fig:randomModelStimDisc}). The relevance of correlations can be
quantified by comparing the discriminability of stimulus pairs in the case
of the full population output and in the case in which noise correlations
are removed by shuffling trials. 

In Fig.\ \ref{fig:randomModelStimDisc}, we visualize, for a recurrent network
model, the interplay of the two effects which affect the influence of correlations: on the one hand, how likely it is that the influence of correlations be favorable for a pair of
stimuli, and, on the other hand, how strong the effect of correlations is for a given pair.
For the first question, we utilize the fact that the noise is predominantly along the diagonal direction. Consequently, the effect of correlations is favorable, if average responses differ in a direction orthogonal to the diagonal. The distribution of the angle between these two directions will therefore reflect the likelihood of beneficial correlations. 
To confirm this picture, and to examine the second question, we plot how the effect of shuffling on the signal-to-noise ratio, $S$, depends on this angle (see Eq.\ (\ref{eq:sigToNoise}) for the definition of $S$).
The network parameters, in this case the variability $\rho$ of the elements of the effective connections, affect both the distribution of angles and the effects on $S$ - the distribution of angles via the distribution of average responses, or the signal correlations, and $S$ by the strength of the noise correlations.
 However, it turns out that the effect of shuffling, averaged across all pairs, is the same in each network. This is due to the fact that signal correlations are on average as strong as noise correlations in all of these networks.
In Section \ref{sec:twoPopModel}, we analyze a simplified network model to better understand some of these effects.

\subsection{Variability and correlations in populations of mouse cortical auditory
 neurons}
 We use the theoretical results presented in the previous sections to
analyze the responses of a population of neurons to different stimuli. The data set
contains the firing rates, collected during a certain time interval, in response to the presentation of different sound
 stimuli in a number of trials (see Section \ref{sec:methodsExp} and \cite{Bathellier2012}).
We will compare the properties of covariances (across trials) and average responses  
with the predictions of our different models. Based on the models, we can then evaluate the effect of network generated variability on stimulus discrimination. 

\begin{figure}[H]
\begin{center}
\includegraphics[width=13.2cm]{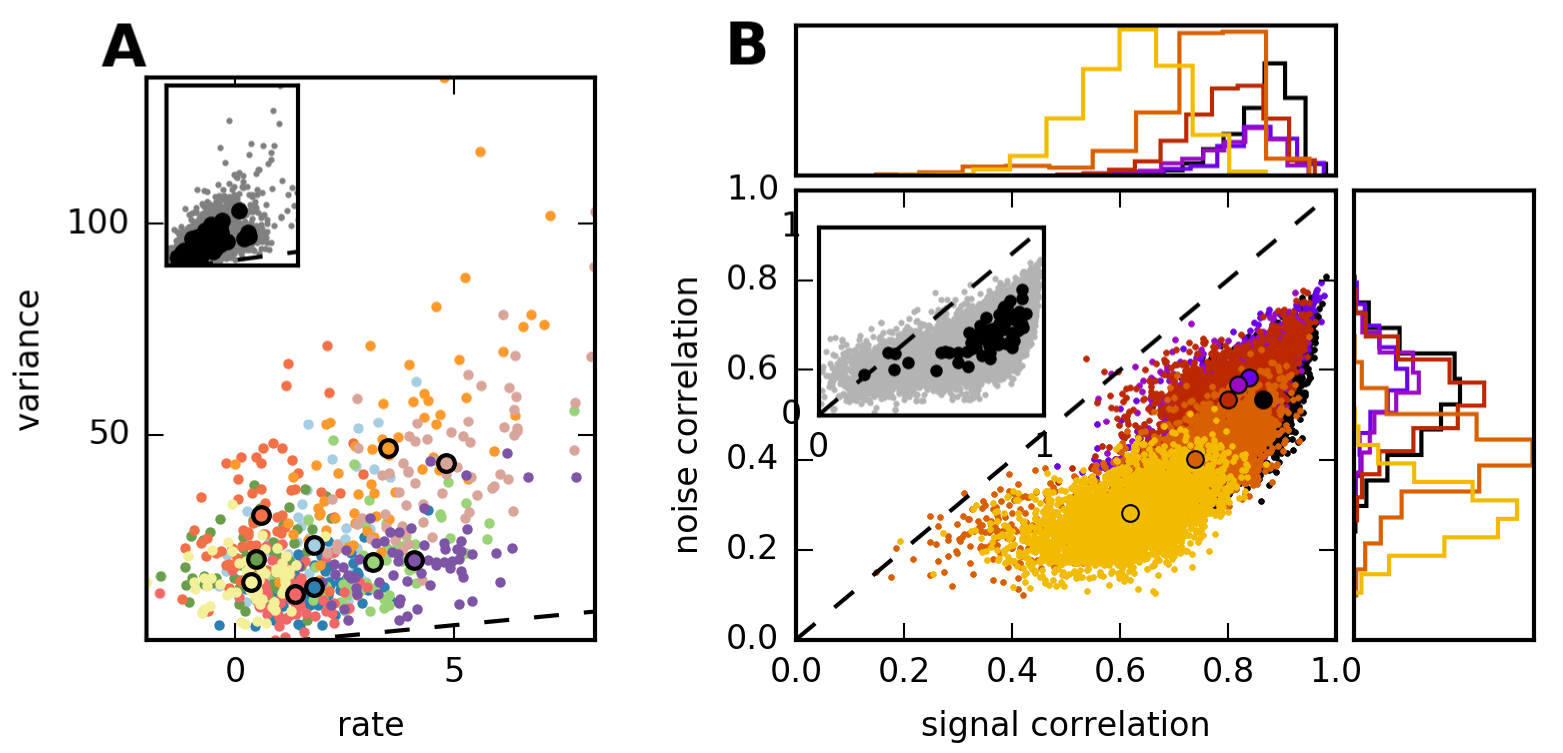}
\end{center}
\caption{ \textbf{Variability and correlation in mouse auditory cortex.} A:
Average response versus response variance in a single population.
Colors correspond to 10 randomly chosen stimuli. Dots correspond to single
neurons, circles to averaged values across population. Dashed line indicates
identity. Inset: All neurons for all stimuli. B: Scatter plot of signal
correlation coefficient (across stimuli) vs.\ noise correlation coefficient
(averaged across stimuli) for different neural populations measured in the
same animal, marginal histograms at top and bottom. Signal and noise
correlations are correlated across pairs. Correlations are high in general,
but the amount of signal and noise correlations varies strongly across
populations. Circles denote average across pairs in each populations. Inset:
Black circles: Average signal and noise correlations in all measured
populations. Grey dots: individual pairs, 5 \% randomly chosen from all
experiments. }
\label{fig:expNoiseCoding}
\end{figure}

The trial-to-trial variability in single-neuron output is large
(supra-Poisson, Fig.\ \ref{fig:expNoiseCoding}). 
Neurons with larger average response exhibit also a larger variability in their 
responses. This tendency is observable across different neurons responding to a given stimulus, for individual neurons across stimuli and for the average population response across stimuli.
The pairwise correlations in trial-to-trial variability are generally high 
and pairs with high noise correlations tend to have strong signal correlations. In other words, because  signal correlations measure the similarity of average responses across different stimuli,  neurons with similar tuning properties also have similar trial-to-trial variability. 
Additionally, in populations with strong average signal
correlations, noise correlations are strong as well. 
These relations between noise and signal correlations can be reproduced
 in the recurrent network model (Fig.\ \ref{fig:randomModelCorrs}). However, similar
results may be possible also in alternative scenarios for the generation of
correlated variability. We compare the different models in the next
section. 

\subsection{Analysis of population response variability in mouse auditory
cortex}
Motivated by our investigation of network models, we first examine the
relation between the shape of the response distributions for each stimulus
and the pattern of average responses for the entire set of stimuli. In Fig.\ \ref%
{fig:expSigs}, we display the dependence of the normalized standard
deviation projected on the direction of the mean response, $\sigma _{\mu
}/\sigma _{all}$, and the normalized standard deviation projected on the
diagonal direction, $\sigma _{d}/\sigma _{all}$, on the angle between average response and diagonal, through $\cos(d, r)$ (see Section \ref{sec:methodsRepVar} for details and definitions).
 We observe a strong dependence on $\cos(d, r)$
of the
variance projected on the direction of the mean response, but not of the
variance projected on the diagonal direction. The dependence of $\sigma
_{\mu }/\sigma _{all}$ on the stimulus is much stronger in nearly all the measured populations. This behavior is consistent with a network model, either feed-forward or recurrent,
 but not with a model with shared gain
fluctuations, where a large part of the variance is 
consistently in the direction of 
the mean response (see Section \ref{sec:theoryAltern}.)
 Note, however,
that in the data firing rates are measured relative to the spontaneous
activity; this is reflected in the presence of negative values in the
stimulus evoked activity. Such an offset could  affect our
comparison because we do not know the true value of the mean response. 
To account for this possibility, we searched for the best possible offset,
 assuming the model of gain fluctuations for the stimulus-dependent covariances. 
We then corrected the firing rates by this offset, and evaluated the stimulus
dependence of $\sigma_{\mu}$ and $\sigma_d$, as before (see Appendix \ref{app:gainFlucTest} for details),
 but found no qualitative change in the results. However, we did not test
for mixed models, and it is possible that part of the correlated
variability can be explained by shared fluctuations.

Based on the results presented above we conclude that, of our three 
scenarios, feed-forward and recurrent network models are more consistent with
the data. In the following, we fit the parameters of these models to the data
 to find which of these two provides a better fit. 
For this comparison, we analyze the dependence of the average variances, $\langle C_{ij} \rangle_{i\neq j}$, and covariances, $\langle C_{ii} \rangle_{i}$, on the population-averaged response, $\langle r\rangle(s)$. 
In the recurrent network (Eqs.\ (\ref{eq:varScalingRec}) and (\ref{eq:covScalingRec})), both variances and covariances increase linearly with $\langle r \rangle$ across stimuli, while in the feed-forward network (Eq.\ (\ref{eq:covScalingGain})), mean covariances are not expected to depend strongly on $\mur$.

\begin{figure}[H]
\begin{center}
\includegraphics[width=19.05cm]{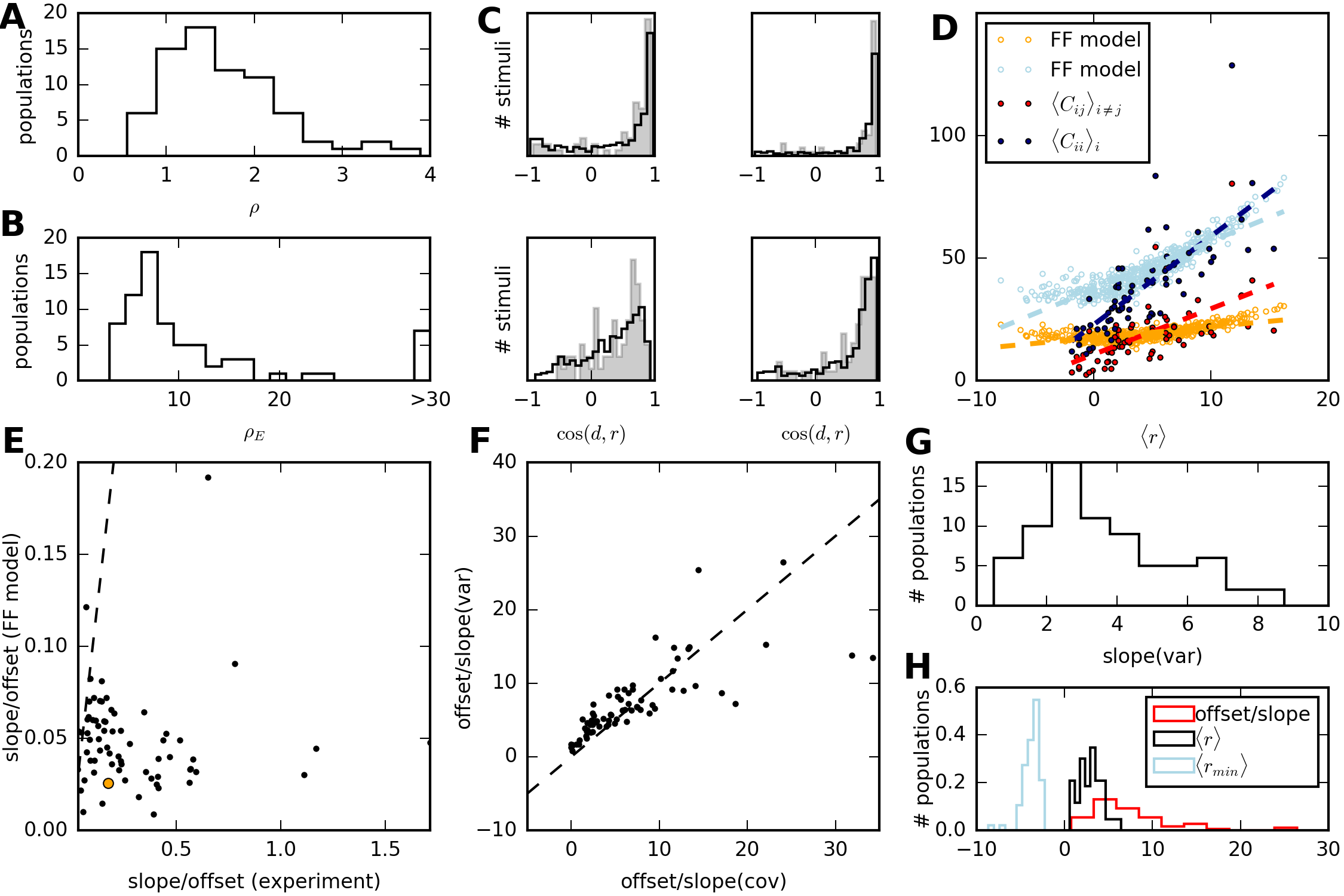}
\end{center}
\caption{ \textbf{Scaling of covariances with average response in experiment and model. } A, B: Distribution of parameters $\rho$ and $\rho_E$ (variability of network connections and stimulus input)
 estimated from all
experiments. C: Histograms of $\cos(d,r(s))$ from mean responses across stimuli,
for a selection of populations measured in one animal. Filled histograms:
experiment. Solid lines: Model results for 500 randomly generated stimuli
and one network with random effective connections, parameters inferred from
data.
D: Population averaged response versus population averaged variances and covariances
for all stimuli (full circles) in one experiment and from corresponding
feed-forward model (empty circles). Dashed lines represent linear fits. E:
Scatter plot of ratio slope/intercept from linear fit in all experiments versus
corresponding value in the feed-forward model. Orange circle indicates
population used in panel D.
F: The ratio intercept/slope is, across experiments (dots), consistent for covariances and variances. G:
Distribution of estimated slopes for average variances. H: Distribution of
estimated $\langle {V_{\mathrm{ext}}} \rangle$ (intercept/slope from fit to variances) in
comparison to average and minimum firing rate across experiments. }
\label{fig:slopeComp}
\end{figure}

To examine whether the experimental data is consistent with the feed-forward model,
 we fitted parameters to the activity of each experimentally observed
population (see Appendix \ref{app:estimation}), and generated a surrogate
data set of responses and covariance matrices, with a matching number of
neurons and stimuli. 
In particular, we obtained values for the variability of the input ensemble, $\rho_E={\var{r_{\mathrm{ext}}}}/{\exval{r_{\mathrm{ext}}}^2}$, and of the network elements, $\rho = {\var{F}}/{\exval{F}^2}$.
It turns out that $\rho_E, \rho>1$: for both input and network elements the variance is much larger than the mean  (see Fig.\ \ref{fig:slopeComp}A-B). The reason is 
the high variability of average responses across stimuli.
With these parameters, the
statistics of the distribution of average responses are well reproduced (Fig.\ \ref{fig:slopeComp}C).
As we discussed in Section \ref{sec:theoryAltern}, for a high value of $\rho_E$ the feed-forward model does not predict a strong relation between $\langle C_{ij} \rangle_{i\neq j}$ and  $\langle r\rangle(s)$. We quantify this relation by the ratio of slope and intercept from a linear fit to the data, which  indicates how strongly the average covariances increase for stimuli
that evoke strong population responses. 
 In the feed-forward scenario, this
behavior does not obtain: the slopes of the linear fits in the model
relative to the intercepts are too low in comparison to what is observed in the
experiment (Fig.\ \ref{fig:slopeComp}D, E). The feed-forward model thus cannot reproduce both the large variability of neural responses across stimuli and the increase of the average covariance with the average response. 

In a recurrent model, this increase is expected, because noise generated by the neurons is propagated through the network: the covariances are proportional to the average response  (see Eq.\
 (\ref{eq:covScalingRec})).  Moreover, it predicts that the  ratios of intercept and slope in the linear behaviors of variances and covariances are the same (it indicates the variance of the external input). 
 Indeed, linear fits to
these relationships reveal approximately consistent ratios of intercept and
slope. The estimated parameters are summarized in Fig.\ \ref{fig:slopeComp}G and
H. The external noise, estimated from the ratio intercept/slope, turns out to be
of the same magnitude as the average rate. Interpreted in terms of the
model, the noise resulting from Poisson spike generation thus contributes as
much to neuron variance as the external input. This combined variability is
 propagated through the network, and, on average, multiplied by a factor $N\langle {B}\rangle
^{2}>1$, corresponding to the slope in Eq.\ (\ref{eq:covScalingRec}). Because this factor is larger than one, both average response and noise are amplified by 
the recurrent connections.

In summary,  we find that a model of recurrently connected neurons with
random effective connections captures the observed activity
statistics, in particular the relations between average response and average covariances as well as the consistent direction of population fluctuations across stimuli. Both purely feed-forward networks and gain fluctuation models are not consistent with all of these observations. We note, however, that our conclusions are based on a relatively small data set. A larger number of trials and measurements of absolute values of firing rates would be desirable for a more stringent test. 

\begin{figure}[H]
\begin{center}
\includegraphics[width=15.2cm]{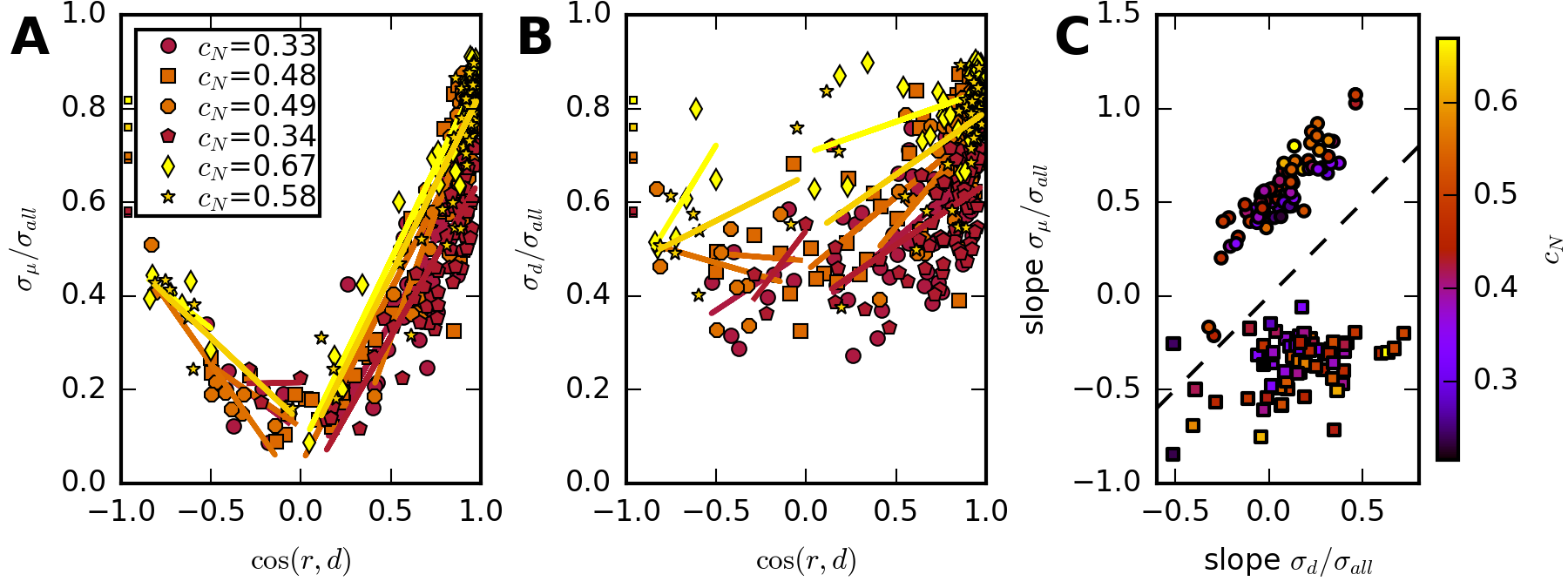}
\end{center}
\caption{ \textbf{Dependence of noise distribution orientation on average response in
data. } A, B: Typical example, 6 neural populations recorded in the same animal.
Relative variances projected on mean and diagonal direction, versus cosine
of angle between mean response and diagonal. Each marker corresponds to a
different stimulus. Different markers/colors denote different populations.
 Squares to the left side indicate $\protect\sqrt{%
\langle {C_{ij}} \rangle_{i\neq j }/\langle {C_{ii}} \rangle_{i} }$. Solid
lines: linear fit.
C: Slopes from linear fits as in A/B from $\protect\sigma_{\protect\mu}/%
\protect\sigma_{all}$ vs.\ slope from $\protect\sigma_{d}/\protect\sigma%
_{all}$ for all measured populations. Circles correspond to slopes for
positive, squares to negative $\cos(r,d)$. Colors in all panels indicate
value of $c_N$ in populations. }
\label{fig:expSigs}
\end{figure}

\subsection{Influence of noise correlations on stimulus discrimination}

\begin{figure}[H]
\begin{center}
\includegraphics[width=13.2cm]{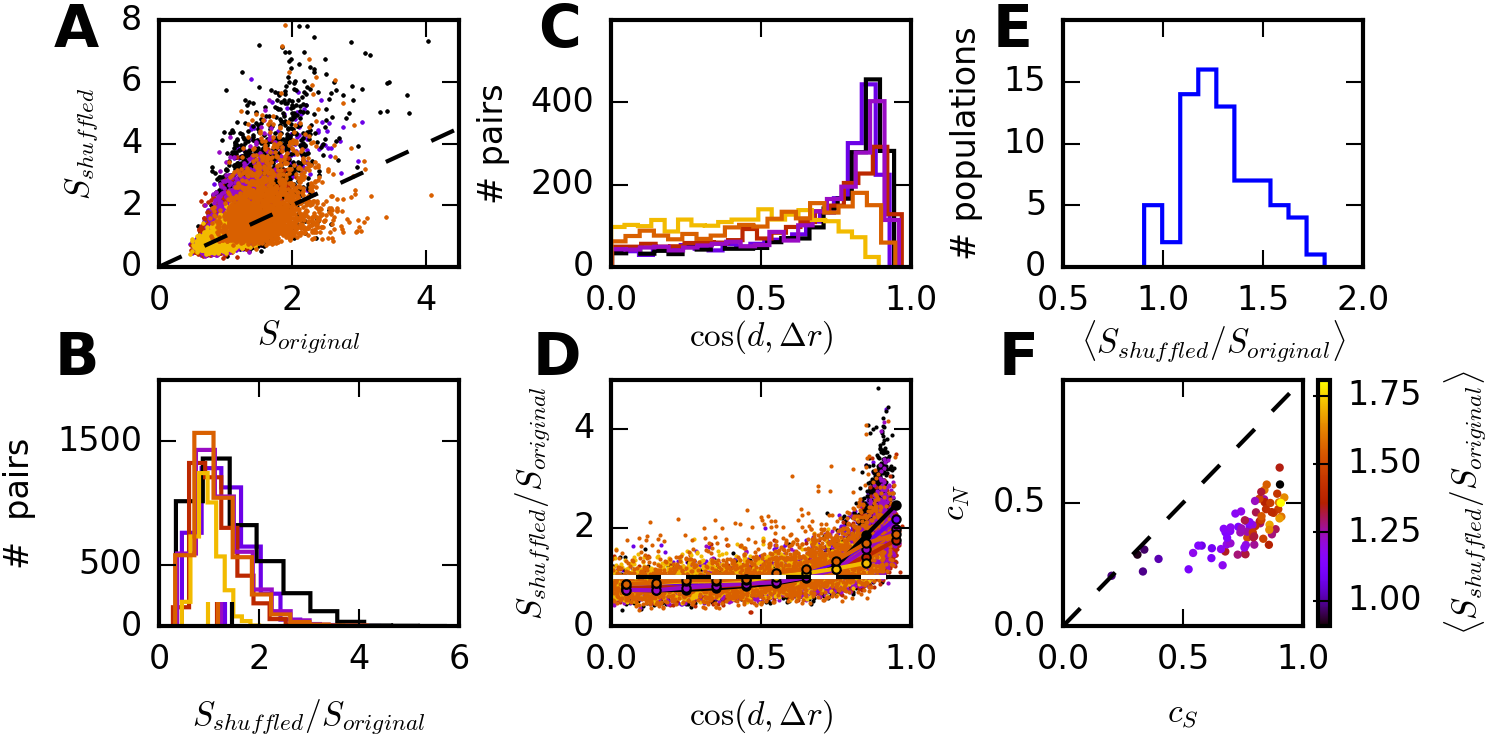}
\end{center}
\caption{ \textbf{Effect of correlations on stimulus discrimination.} A:
Signal-to-noise ratio $S$, dots correspond to pairs of stimuli. Value from
observed covariance matrix versus the one based on shuffled trials, for
different populations in the same animal.
B: Same data, distribution of $S(\mathrm{shuffled})/S(\mathrm{original})$.
Broad distribution, but mean value greater one for all populations (short
solid lines) indicates that shuffling increases information on average. C:
Distribution of cosine of angle between diagonal and average response
difference across stimulus pairs for neural populations. Large cosines/small
angles are most frequent. D: Scatter plot of effect of correlations on
discrimination. Small dots correspond to stimulus pairs. Connected large
dots indicate average value in a bin centered at the corresponding location
on the x-axes. E: The distribution of mean values across all measured
populations shows that correlations are on average unfavorable for almost
 all populations. F: Average
effect of correlations is stronger for increasing signal correlations. }
\label{fig:codingShuff}
\end{figure}

We quantify the influence of noise correlation on stimulus discrimination by
the ratio $S_{\text{shuffled}}/S_{\text{original}}$. $S$ is calculated for the data set before and after shuffling trials. It is defined in Eq.\ (\ref{eq:sigToNoise}) and denotes, for a pair of stimuli, the difference in average response divided by the standard deviation of the responses, both projected on the most discriminative direction. Larger values indicate that stimuli are easier to discriminate. If $S_{\text{shuffled}}/S_{\text{original}}$ is larger than one, then removing correlations by shuffling trials improves stimulus discrimination.
 Across pairs of stimuli, this signal-to-noise ratio varies
strongly (Fig.\ \ref{fig:codingShuff}). On average, stimuli are slightly
easier to discriminate in the shuffled data, so that noise correlations are
weakly unfavorable to the encoding of the stimulus set used in the
experiments.

A closer analysis of the response distributions reveals that, to a large
degree, the effect of shuffling can be explained by the relative locations
of the two mean responses and the diagonal, as measured by $\cos
(d,r(s_{1})-r(s_{2}))$ (Fig.\ \ref{fig:randomModelStimDisc}). This is because, due to the noise correlations, the main direction of variability 
is along the diagonal.
The overall
effect of shuffling on stimulus discrimination depends on the relation
between noise and signal correlations:
stronger signal correlations lead to a 
stronger dominance of angles that are unfavorable if noise is aligned along the diagonal direction, and in this case shuffling correlations away will benefit stimulus discrimination. 

\subsection{Intuition from a two-population model}
\label{sec:twoPopModel}
The relation between network properties and the statistics of variability,
on the one hand, and the relation between the statistics of variability and
the encoding of stimuli, on the other hand, have been studied widely. Above,
we describe the structures of correlations that emerge from different network
architecture. We also analyze the structure of correlations in data from
mouse cortex, and we relate them to stimulus encoding. Here, we study a
highly idealized recurrent cortical network in order to develop an
understanding of the relation between network characteristics, response statistics, and stimulus encoding. 

\begin{figure}[H]
\begin{center}
\includegraphics[width=19.05cm]{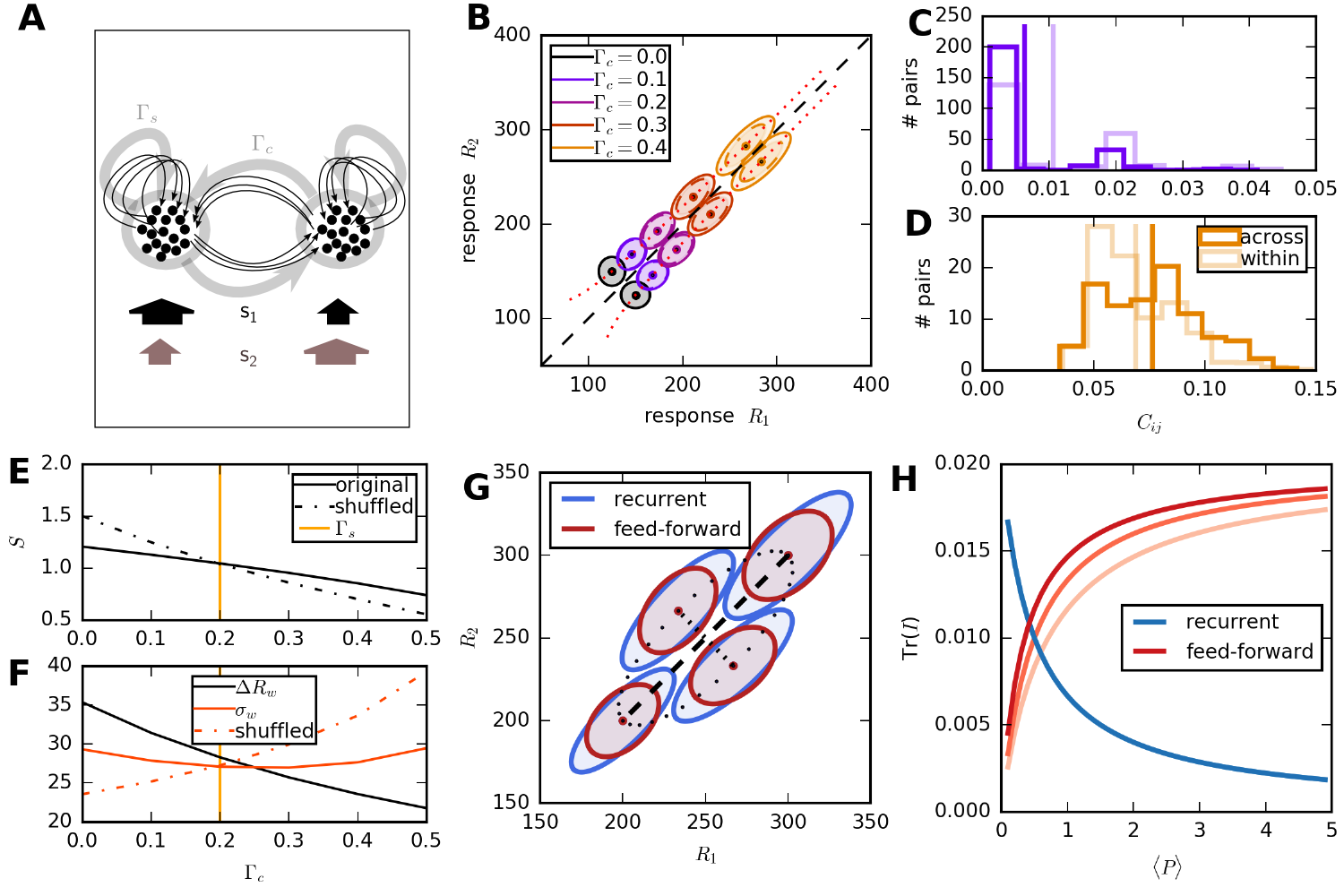}
\end{center}
\caption{ \textbf{Model of two coupled excitatory populations.} A: Circuit
model. Neurons are coupled with excitatory connections within and across two
populations. For each stimulus, one population receives stronger input. Population responses are described by macroscopic variables.
B: Responses to two stimuli (colored dots on dotted red line) are amplified
for stronger cross-coupling (brighter colors). Simultaneously, response
fluctuations are more correlated. If internally generated noise does not
increase with rate, variability is smaller (dashed ellipses).
C and D: Distribution of pairwise covariances within (dark bars) and across
populations (light bars), vertical lines indicate average. If cross-coupling
is weak (C), average covariance is larger within populations than across, if
it is strong (D), covariances between neurons of different populations are
larger.
E: Dependence of stimulus discriminability $S$ on $\Gamma_c$. Strong
couplings implies smaller discriminability and even more so for shuffled
trials. 
F: Increasing cross-coupling reduces difference in mean response (black
line). Effect on variance in the relevant direction differs for correlated
populations (solid orange line)
and model with independent neurons/shuffled correlations (dash-dotted orange
line).
G: Red dots: average response for stimulus ensemble. Ellipses: Response
distributions for four stimuli for the recurrent network (blue) and a
feed-forward network with identical average responses (red). Dotted/dashed
lines: stimulus pairs where correlations have positive/negative influence.%
 H: Linear Fisher information, dependence on average of population
transfer matrix elements (horizontal axes) for recurrent and feed-forward scenario.
Darker lines indicate larger variance of transfer matrix elements. Lines
overlap in the recurrent scenario. }
\label{fig:twoPopulation}
\end{figure}

To understand the effect of recurrent connections on the discriminability of a pair of stimuli, we consider two populations of excitatory neurons. The neurons in each of these populations prefer one of the stimuli, and we want to understand the effect of connections between these differently tuned neurons.
The connectivity is regular, in the sense that each neuron projects to a fixed number of neurons in its own population and to a fixed number of neurons in the other population. Neurons in each population also receive the same external input. We start by considering two stimuli, $s_{1,2}$, where for each stimulus one of the populations receives a stronger input. To discriminate between these stimuli, it is sufficient to consider the network response at the population level, that is, the sum of the response of neurons in each of the two populations. This follows from the regularity of the connections. The distribution of responses across trials  can thus be described by a two-dimensional vector, $R$, of the average population response and a covariance matrix, $\Sigma$, of the variability. 
The sum of the couplings within a population, $\Gamma_s$, and across populations, $\Gamma_c$, determine the response via a symmetric $2 \times 2$ coupling matrix, $\Gamma$. Here, we will manipulate the entries of $\Gamma$, rather than the effective population coupling matrix, $\mathrm{P}=(\id - \Gamma)^{-1}$ (see Fig.\ \ref{fig:twoPopulation} and Methods for further details of the model). 

We start by examining how cross-coupling between the populations can induce beneficial correlations for stimulus discrimination, by considering the effect of connections between the populations on the covariances.
If we introduce more connections across populations, the cross-coupling $\Gamma _{c}$ increases and the response distributions evoked by the two stimuli change (Fig.\ \ref{fig:twoPopulation}).
 The
responses of the two populations are amplified, but also more similar - the
difference in population response, $|R(s_{1})-R(s_{2})|$, decreases.
Simultaneously, the correlation
between the population responses increases, rendering the distributions more
elongated. To evaluate the effect of shuffling trials, we need to consider
the covariances between individual neuron pairs. As we show in detail in 
Appendix \ref{app:popModel}, covariances reduce the overlap of the distributions if $\Gamma_c>\Gamma_s$. Then,  covariances between neurons belonging to different populations are larger, on average, than for neurons of the same sub-population, and the signal-to-noise ratio, $S$, becomes smaller if
covariances are shuffled away (Fig.\ \ref{fig:twoPopulation}C, D).
While it thus appears that strong coupling between differently tuned neurons 
induces beneficial pairwise correlations this is not always the case. 

To see this, we compare networks with different parameters, and disentangle the effects of cross-connections on average responses, variances, and covariances.
Increasing the coupling between populations  decreases  $S$  both in the shuffled and in the
original cases (Fig.\ \ref{fig:twoPopulation}E), but this suppression is stronger when covariances are
shuffled. The suppression can be traced back to the amount of noise produced through spike generation. Recurrent amplification increases
firing rates, and the variance of Poisson neurons increases with rate.
If the amount of noise remained constant,
 $S$ would not depend on $\Gamma_c$: larger coupling induces stronger correlations which reshape noise along a direction  irrelevant for discrimination, but at the same time average responses approach each other. These two effects precisely cancel each other, ultimately because a linear transformation of responses by the network cannot reduce noise. 

Above, we only considered a pair of stimuli for which beneficial correlations are realized in our network. We now evaluate the effect of cross-connections on a more general stimulus ensemble.
We define it via the input vectors for the two populations, $(1+\Delta\cos(\phi),1+\Delta\sin(\phi))^T$ for  $\phi \in (0,2\pi)$ and a constant $\Delta$ (see Fig.\ \ref{fig:twoPopulation}G and Appendix \ref{app:popCorrs}).
For this ensemble we see that inevitably correlations will be unfavorable for certain stimulus pairs, depending on the stimulus dimension in which their inputs differ. The averaged effect of shuffling across this two-dimensional stimulus ensemble depends not only on the noise correlations, but also on the distribution of the input, which affects the distribution of the average responses, i.e.\ the signal correlations. 

If excitatory recurrent networks amplify internally generated noise, are
feed-forward architectures generally more efficient? In the following,
we compare the performances in the two scenarios.
As we noted before, the
difference between the two scenarios is that the noise produced by spike
generation in the recurrent network is fed back into the network, and
therefore correlated, while it is independent in the feed-forward network.
The measure of performance we used so far is the discriminability of
pairs of stimuli. We want to compare the effect on information of a recurrent network, with
a transfer matrix $\mathrm{P}$ defined by the connectivity parameters, to a
feed-forward network with identical transfer matrix, for changes of
a high-dimensional input in arbitrary directions.
 A quantity that measures the information of a network response statistics
in the case of a continuous stimulus is the linear Fisher information matrix.
In the two-population model, it is given by 
\begin{equation}  \label{eq:FIrec}
I_R=(D[R]+\Sigma_{\mathrm{ext}})^{-1}
\end{equation}
in the recurrent case and by 
\begin{equation}  \label{eq:FIff}
I_F=\mathrm{P}^{T}(\mathrm{P}\Sigma_{\mathrm{ext}}\mathrm{P}^{T}+D[R])^{-1}\mathrm{P}
\end{equation}
in the feed-forward case. Here, $\Sigma_{\mathrm{ext}}$ denotes the variance of the external input to the two populations (see Appendix \ref{app:secLFI} for details).  
As figures of merit for the coding of high-dimensional stimuli, we use
the traces of IR and IF, for the recurrent and feed-forward cases
respectively. In the recurrent network, this quantity depends
only on the firing rates. These, in turn, depend on the average strength of the effective connections, $\langle P\rangle$  (Fig.\ \ref{fig:twoPopulation}H): the larger $\langle P \rangle$, and therefore the
response, the more noise is generated, and the more information is lost. In
a corresponding feed-forward network, information depends also on the
correlations. However, because the network-generated noise is uncorrelated,
increasing firing rates effectively improves the signal-to-noise ratio, and
therefore increases the information. Whether a feed-forward or a recurrent
architecture is preferable thus depends on the output firing rate. From this model,
it is not clear how firing rates should be chosen, but additional constraints,
 for example imposed by energy consumption are conceivable.

A simple model with two coupled, uniform populations illustrates how variations in the
network parameters can affect stimulus discriminability, through modulations in the
average responses and variability. Broadly speaking, stimulus discriminability depends on
the interplay between noise creation, average response, and noise correlation, and the relative behavior of these quantities depends on the network architecture.

\section{Discussion}

Despite large amounts of work on the topic, the interpretation of the
observed response fluctuations in population activity, in the context of
computation by neural circuits and information coding, is not elucidated 
\cite{Doiron2016}. In the present study, we seek to understand the influence
of correlated variability on stimulus representation, based on the
exploration of model networks of stochastic neurons together with data
analysis. We modeled neural activity using a set of coupled Poisson
processes in order to study the relations between covariances and average
responses in a neural population, across different stimuli. We found that
these relations differ depending on how covariances are generated, and that
correlations observed in mouse auditory cortex are consistent with those
generated in a recurrent network model. This model also helps us interpret
the effect of the observed correlations on the discriminability of stimuli
and the nature of the propagation of information through the network.

\subsection{Origins of the structure of correlations in neural circuits}

In the examination of the impact of correlated response variability on
stimulus representation, it is useful to be aware of the possible range of
its mechanistic origin. Spike train variability has been attributed to the
dynamics in recurrent networks \cite{VanVreeswijk1998,Brunel2000}. Along
with this source of variability, in stable states of asynchronous activity,
recurrent network models can also produce transitions between different
states of global activity \cite{Litwin-Kumar2012, Ponce-alvarez2013,
Ostojic2014}. Alternatively, external signals fed to sub-populations of
neurons, such as gain fluctuations induced by upstream areas, were pointed
to as causing high variability and correlations \cite%
{Goris2014,Ecker2014,Lin2015,Scholvinck2015,Franke2015}.
 Population-wide gain fluctuations can also reproduce dependencies between noise
 and signal correlations \cite{Ecker2015}.

In general, it is difficult to differentiate shared input or gain
fluctuations, and recurrent connections, as the origin of correlated
variability, based on measured activity. Here, instead of focusing on
temporal dynamics \cite{Paninski2004a,Pillow2008,Macke2011}, we approach the
problem by considering the relation between the network architecture and the
structure of population activity statistics, when a neural population is
presented with an ensemble of stimuli. Our analysis makes the assumption
that noise resulting from spike generation in neurons can be modeled as a
Poisson process, and, thus, depends on the firing rate. If the noise
generated within the network is purely additive and independent of firing
rates, it is indistinguishable from external noise that is simply filtered
by the network \cite{Grytskyy2013}. In the Poisson model, however, the
interplay of internally generated noise and variability due to external
signals imprints its signature on the dependence of covariances on firing
rates.

\subsection{Influence of correlations on stimulus discrimination in neural
populations}

We use a recurrent network with random effective connections to estimate the
influence of the variability generated by the network dynamics on stimulus
representation. Based on the paradigm that response variability amounts to
noise that downstream neurons have to cope with, a number of authors have
argued that noise correlations can benefit information coding \cite%
{Averbeck2006a,Ecker2011,Hu2014,DaSilveira2014,Franke2015,Zylberberg2016}.
Furthermore, recent studies have examined specifically the impact of
correlations generated by recurrent networks. Reference \cite%
{Moreno-bote2014} stresses that it is the structure of the noise, more than
its magnitude, which determines the role it plays for information coding.
Harmful structures of noise correlation, causing information
to saturate in large neural populations, are the consequence of the amount of
irreducible external noise and potentially sub-optimal processing \cite{Kanitscheider2015a},
 but these structures
 are difficult to pinpoint in detail in measured activity \cite{Moreno-bote2014}. 
By contrast, it was demonstrated that noise internally generated in a network
 via spike generation
 can be averaged out in feed-forward or recurrent networks, depending on the
connectivity in the network and the redundancy of the population code \cite%
{Beck2011}.

In this work, the goal was to start from the observed properties of the
activity, and to interpret mean activity and noise correlations based on
outcomes of model network dynamics. 
Variations of network properties, such as the architecture of connections, change both the
mean activity and noise correlations, both of which influence the accuracy of the neural code.
Consequently, it is useful to consider both simultaneously to  understand
network effects.
 The effect of recurrent dynamics on input information depends
mostly on the amount of internally generated noise. Our toy model, as well
as Ref.\ \cite{Beck2011}, argue that recurrent amplification suppresses
information. 
The reason is essentially that, for Poisson neurons, larger firing
rates imply a higher variance. The stronger the recurrent connections, the more the average response is amplified, and the more noise is produced internally. Because this noise is fed back into the network and thus amplified furthermore, the variance is increased to the point that the signal-to-noise ratio decreases.  
We also show that under simplifying assumptions, including a random effective network
 architecture and a large, random 
stimulus ensemble, the amplification factor of the network and the amount of noise
can be estimated from measured responses.

Limitations of our analysis result, in particular, from the assumption of
linearly interacting Poisson processes. Nonlinear transfer functions can
affect the relation between rates and correlation in individual pairs \cite%
{DelaRocha2007,Doiron2016}. Using Poisson neurons also implies that the
generation of noise in the network is attributed to individual neurons. The
tractability of the model, however, allows us to obtain relations between
observable statistics, like noise and signal correlations, which can be used
as a starting point to interpret correlated variability in terms of its
mechanistic origins.

\subsection{Variability in mouse auditory cortex}

We tested the consistency of different scenarios involving interacting
Poisson neurons with measured neural population responses, and we
interpreted the observed variability in terms of a recurrent network model.
The measured population activity is given as spike counts in 200 ms bins, so
that correlations are defined over a relatively slow timescale and,
therefore, do not take into account fast temporal structure (as noted in,
e.g., Ref.\ \cite{Luczak2013}). Both experimental variability and
correlations are high \cite{Cohen2011}, so that the effects of correlations
are potentially strong and amenable to an analysis such as ours. Although
variability and correlations are often high in anesthetized animals \cite%
{Lin2015}, noise and signal correlations were somewhat weaker in comparable
experiments with anesthetized animals \cite{Rothschild2010}. 
It is conceivable that part of the correlations are due to
experimental artifacts from the calcium imaging technique used in the experiments (like scattering of fluorescent light by the neuropil), but intracellular recordings of rates were
consistent with calcium recordings \cite{Bathellier2012}. 

We model the set of stimuli as a generic, high-dimensional ensemble, and the
network connections at the level of a transfer matrix. When analyzing the
discriminability of pairs of stimuli, this enables a broader analysis than
one relying on a single stimulus dimension or average correlation. The
recurrent network model explains not only the scaling of variances and
covariances with firing rates, but also the relation between signal and
noise correlations, and the effect of shuffling away covariances on stimulus
discrimination, in a natural way. Differences among populations of neurons
can be attributed to changes in the effective connectivity within
populations.

Recent studies have explained correlated variability in terms of ``external''
gain fluctuations acting on a (sub-)population \cite%
{Goris2014,Ecker2014,Lin2015,Scholvinck2015, Franke2015}.
 We showed that recurrent networks of strongly connected
neurons can exhibit population-wide fluctuations as well.   Estimated
model parameters indicate an amplification of input signals by excitatory
recurrent connections. This suggests that the noise that can be traced back
to spike generation in the population --- noise harmful to coding --- is
comparable in magnitude to the external input noise.

Our data analysis suffers from an unknown offset in the firing rates, a
relatively small number of trials, and the low temporal resolution. We
therefore only compared different prototypical scenarios without trying to
evaluate the outcome of a mixed model. In experimental populations of
neurons, the statistics of activity can obviously result from a combination
of both feed-forward and recurrent processes.

\section{Acknowledgments}
We thank Brice Bathellier for providing the experimental data set as well as
commenting on it, and for discussions.

This work was supported by the CNRS through UMR 8559
and the SNSF Sinergia Project CRSII3\_141801.

\section{Author Contributions}
Conceived and designed the experiments: VP RS. Performed
the experiments: VP. Analyzed the data: VP. Contributed
reagents/materials/analysis tools: VP RS. Wrote the paper: VP RS.

\newpage

\section{Appendix}
\subsection{Numerical simulations: details and parameters}

\label{app:figDetails} In Fig.\ \ref{fig:randomModelSigs}, covariances  were calculated for recurrent networks from Eq.\ (\ref{HawkesCovs}), with the addition of a rate offset $a=4$, so that $C=(\mathbb{I}-G)^{-1}\bigl(D[r]+a+D[V_{\mathrm{ext}}]\bigr)(\mathbb{I}-G^{T})^{-1}$. The
stimulus ensemble consisted of $200$ input vectors $r_{\mathrm{ext}}$ of size $N=60$, with
entries independently chosen from a normal distribution with mean $0.2$ and
standard deviation $1$. Rates $r$ were calculated as in Eq.\ (\ref{HawkesRates}), and we assumed Poisson input with $V_{\mathrm{ext}}=|r_{\mathrm{ext}}|$. 
For the  networks, four
connectivity matrices $G$ of size $N=60$ were generated. Their entries were
chosen from a normal distribution with standard deviation $1.5/N$ and mean $%
0.8/N-0.9/N$, respectively. 
For the analytic predictions from Eqs.\ (\ref%
{eq:covScalingRec})-(\ref{eq:covScalingGain}), the mean and variance of the
transfer matrix $B=(\mathbb{I}-G)^{-1}$ were calculated numerically. 

For Fig.\ \ref{fig:covMeanDep}, for the input ensemble for the recurrent network, elements of $200$ random input vectors $r_{\mathrm{ext}}$ of
length $N=60$ were chosen from a normal distribution with mean 0.4 and
variance 2. Connectivity in the recurrent network was as in Fig.\ \ref%
{fig:randomModelSigs}. For rates and covariances of the feed-forward scenario,  we use Eqs.\ (\ref{FFNrates}) and (\ref{covsFFN}). 
We set $F=10\cdot B$,
and scaled the input vectors by a factor $1/10$ to obtain rates and
correlations of comparable size as in the recurrent network. Negative output rates were set to zero. For the gain
fluctuation model, responses of the network with $\rho = 1.4$ were used as the ensemble of stimulus responses. From these, covariances were calculated, see Eq.\ (\ref{FF1covs}), for which the variance of external fluctuations $V_{\mathrm{ext}}$ was varied. For this figure, no rate offset was used, $a=0$.

The networks in Fig.\ \ref{fig:randomModelCorrs}C were identical to the ones
used above. 50 stimuli $r_{\mathrm{ext}}$ with random entries were generated with mean 0.2 and
standard deviation 1.5. Inputs for different neurons were correlated with
correlation coefficient $c_{\mathrm{in}}=0.05$ across stimuli. Rate offset was 4.

In Fig.\ \ref{fig:randomModelStimDisc}, networks and stimulus parameters were the same
as in Fig.\ \ref{fig:randomModelCorrs}, but there were no input signal
correlations, $c_{\mathrm{in}}=0$.

In Fig.\ \ref{fig:twoPopulation} the network connects two populations
of 100 neurons each. For the pair of stimuli, panel B, neurons in the population that
was more strongly excited received an input of $1+\Delta$, with $\Delta=0.2$,
and the others input $1$. More general inputs (panel G) were defined by setting the inputs $R_{\mathrm{ext}}=(1+\cos(\phi)\Delta,1+\sin(\phi)\Delta)$ for $\phi\in(0,2\pi)$. Each neuron had a fixed number of $n_{EE}$ postsynaptic partners within the
same populations.
Coupling strength between connected neurons was set to $g=0.01$. We chose
a population coupling within populations of $\Gamma_s=0.2$, and the
number of connections between neurons within populations such that $\Gamma_s=n_{EE}g$.
 Connections across populations were chosen accordingly to
realize across coupling values $\Gamma_c=0,\dots,0.4$. To visualize the
effect of the increase in internally generated noise due to increased rates (panel B),
we compared the covariances $\Sigma=B^pD[R](B^p)^T$ for increasing $\Gamma_c$
to the ones where in $D[R]$ the rates of a network with $\Gamma_c=0$ were
used.

In panels E and F, we calculated $S_{original}$ by inserting the population covariances and responses in Eq.\ (\ref{eq:sigToNoise}) for the two stimuli, for the correlated case. For the shuffled case, the full matrix covariance matrix $C$ was generated and the
off-diagonal elements set to 0, such that 
\begin{equation}
\Sigma_{shuffled}= 
\begin{pmatrix}
\sum_{i\in E} C_{ii} & 0 \\ 
0 & \sum_{i\in E^{\prime }}C_{ii} \\ 
\end{pmatrix}.%
\end{equation}

For panels E-H, population level parameters in the matrices $\Gamma$ or $P$ were varied directly. We set the external noise $\Sigma_{\mathrm{ext}}=ND[R_{\mathrm{ext}}]$.
 In panel G, 
$\Gamma_c=0.4$ and $\Gamma_s=0.2$. In panel H, we fixed for the population transfer matrix $P$ the ratio $\rho=\langle {P} \rangle^2/\mathrm{var}({P})$ to the values $\rho=0.5,1,2$. For each of these values, 
$\langle {P} \rangle$ was varied between $0.1\dots 5$, so that $\mathrm{var}({P})$ was fixed. From $\langle {P} \rangle$ and $\mathrm{var}({P})$, we calculated
$P_s$ and $P_c$, see
Appendix \ref{app:popCorrs}.

\subsection{Common framework for the generation of correlations}

\label{app:commonFramework} Here we formulate the three model scenarios
described in Sections \ref{methods:hawkes} and \ref{methods:alternative} as
special cases of the interacting point processes framework defined in Eq.\ (%
\ref{HawkesDynamics}). The derivation is similar as in \cite%
{Hawkes1971,Beck2011}. We consider an extended network of two populations
which receive only constant input $r_{\mathrm{full}}(t)\equiv r_{\mathrm{full}}$. Neurons are divided into an observed
population $O$ and an unobserved external population $U$. The external
population projects to the observed population, but does not receive
feedback. The coupling matrix of integrated kernels of the full system is a block matrix of the
shape 
\begin{equation}
G_{\mathrm{full}}=%
\begin{pmatrix}
E & 0 \\ 
F & G \\ 
\end{pmatrix}%
.
\end{equation}
The coupling between nodes within the external network is described by the matrix $E$.
If $E=0$, external inputs are independent Poisson processes. The
feed-forward weights to the observed network are defined by $F$, and by $G$
the recurrent connections within the observed population. The input vector
to the system is $r_{\mathrm{full}}=(r_0,0)$. The components of the vector $r_{0}$
together with $E$ determine the firing rates of the input nodes. The $0$ represents a vector of zeros, so that there is no constant input directly to the observed population.

The time dependent firing rates of neurons $\ k\in U$ in the external input population are
determined by 
\begin{equation}
\tilde r_k(t)=\sum_{l\in U}\int_0^{\infty} \tilde e_{kl}(\tau)\tilde
s_l(t-\tau)d\tau+r_{0,k}
\end{equation}
and the rates of neurons $\ i\in O$ in the observed population by 
\begin{equation}
\tilde r_i(t)=\sum_{j\in O}\int_0^{\infty} \tilde g_{ij}(\tau)\tilde
s_j(t-\tau)d\tau+\sum_{k\in U}\int_0^{\infty} \tilde f_{ik}(\tau)\tilde
s_k(t-\tau)d\tau.
\end{equation}

The transfer matrix of the full system is 
\begin{equation}
B_{\mathrm{full}}=(\mathbb{I}-G_{\mathrm{full}})^{-1}=%
\begin{pmatrix}
(\mathbb{I}-E)^{-1} & 0 \\ 
(\mathbb{I}-G)^{-1}F(\mathbb{I}-E)^{-1} & (\mathbb{I}-G)^{-1}%
\end{pmatrix}
\equiv 
\begin{pmatrix}
B_E & 0 \\ 
BFB_E & B.%
\end{pmatrix}%
\end{equation}
The average rates of the system are,  from applying Eq.\ (\ref{HawkesRates}) to the full system, 
\begin{equation}
(\mathbb{I}-G_{\mathrm{full}})^{-1}r_{\mathrm{full}}=B_{\mathrm{full}}r_{\mathrm{full}}=(B_Er_0,BFB_Er_0)^T\equiv (r_{\mathrm{ext}},r)^T,
\end{equation}
where $r_{\mathrm{ext}}=B_Er_0$ are the rates of the input neurons and $r=BFB_Er_0=BFr_{\mathrm{ext}}$
correspondingly, the rates of the observed neurons. From Eq.\ (\ref{HawkesCovs}), the covariance matrix is
given by 
\begin{equation}
C_{\mathrm{full}}= 
\begin{pmatrix}
B_ED[r_{\mathrm{ext}}]B_E^T & B_ED[r_{\mathrm{ext}}]B_E^TF^TB^T \\ 
BFB_ED[r_{\mathrm{ext}}]B_E^T & B\Bigl(D[r]+FB_ED[r_{\mathrm{ext}}]B_E^TF^T \Bigr)B^T%
\end{pmatrix}%
.
\end{equation}
The block $C_E\equiv B_ED[r_{\mathrm{ext}}]B_E^T$ at the upper left describes the
covariances of the external population. The block at the lower right, 
\begin{equation}  \label{generalCovs}
C\equiv B(D[r]+FC_EF^T)B^T
\end{equation}
describes how covariances in the observed network depend on the properties
of the unobserved input neurons.

The recurrent network scenario is obtained for an input population and an
output population of equal size $N$, with $F=\mathbb{I}$ and any diagonal $E$. In this case, $C_E$ is diagonal, and we can identify the elements on the diagonal of $C_E$ with $V_{\mathrm{ext}}$. 

For the feed-forward scenario set $G=0$ and again identify the diagonal
elements of $C_E$ with $V_{\mathrm{ext}}$.

The gain fluctuation model is obtained for $G=0$, and if the input
population consists of a single neuron. The matrix $F$ then corresponds to a
single column vector, and $C=D[Fr_{\mathrm{ext}}]+FC_EF^T$. With $r=Fr_{\mathrm{ext}}$ and setting $%
V_{\mathrm{ext}}=C_E/r_{\mathrm{ext}}^2$ ($r_{\mathrm{ext}}$ is a number in this case), $C=D[r]+rr^TV_{\mathrm{ext}}$.

\subsection{Further details on statistics in different model scenarios}

\label{app:ThreeModels}

\subsubsection{Gain fluctuation model}

We derive relations between covariances and average response in the gain
fluctuation model, in particular the scaling of average covariances with
average rates and the orientation of the response distribution measured by
the projections of the variances in different directions. From Eq.\ (\ref{FF1covs}), pairwise covariances are directly related to average responses, 
\begin{equation}
C_{ij}=\delta_{ij}r_{i}+r_{i}r_{j}V_{\mathrm{ext}}.
\end{equation}
The relation between average covariance and the population averaged response, Eq.\ (\ref{eq:covScalingGain}),
follows from
\begin{equation}
\langle {C_{ij}} \rangle_{i\neq j }=\frac{1}{N(N-1)}\sum_{i\neq
j}r_ir_jV_{\mathrm{ext}}\approx \frac{V_{\mathrm{ext}}}{N^2}\Bigl(\sum_ir_i\Bigr)\Bigl(\sum_jr_j\Bigr)%
=V_{\mathrm{ext}}\langle {r} \rangle^2,
\end{equation}
where the terms $r_i^2$ can be neglected if $N$ is large. 

To measure changes of the response distribution across stimuli, we use the variance projected in the direction of mean response and diagonal direction, $\sigma_{\mu}^2$ and $\sigma_d^2$, respectively. In the following, we give an argument that in this model $\sigma_{\mu}^2$ does not strongly depend on the stimulus, while $\sigma_d^2$ does.

Both quantities are normalized by the sum of the variances 
\begin{equation}
\sigma^2_{all}=\sum_iC_{ii}=\sum_{i}(r_i^2V_{\mathrm{ext}}+r_i)=|r|^2V_{\mathrm{ext}}+N\langle {r}
\rangle.
\end{equation}
The variance projected in the diagonal direction $\bar d=(1,\dots,1)^T/%
\sqrt{N}$ is 
\begin{equation}
\sigma_{d}^2=\sum_{ij}C_{ij}(s)\bar d_i \bar d_j=\sum_{ij}\Bigl(%
r_ir_jV_{\mathrm{ext}}+\delta_{ij}r_i\Bigr)/N=NV_{\mathrm{ext}}\langle {r} \rangle^2+\langle {r}
\rangle.
\end{equation}
To see that there is a strong dependence of $\sigma_{d}^2/\sigma_{all}^2$
on stimulus direction, note that 
\begin{equation}
\cos(d,r)=\bar d \bar r^T=\sum_{i=1}^N\frac{1}{\sqrt{N}}\frac{r_i}{|r|}=%
\sqrt{N}\langle {r} \rangle/|r|.
\end{equation}
For non-vanishing $r$, it follows from $\cos(d,r)=0$ that $\langle {r}
\rangle=0$ and in this case $\sigma_{d}^2/\sigma_{all}^2=0$.

If $\cos(d,r)=1$, $r\propto d$ and $|r|^2=N\langle {r} \rangle^2$, and so $%
\sigma_{d}^2/\sigma_{all}^2=\frac{NV_{\mathrm{ext}}\langle {r} \rangle^2+\langle {r}
\rangle}{NV_{\mathrm{ext}}\langle {r} \rangle^2+N\langle {r} \rangle}$. This value is of
the order of one, if the ratio of $V_{\mathrm{ext}}\langle {r} \rangle^2$ and $\langle {r}
\rangle$, which is of the order of the average noise correlation
coefficient, is not very small (in comparison to one). Consequently, if noise correlations are not too small, the normalized variance
projected on the diagonal direction strongly depends on the direction of the
response vector.

The variance projected on the mean direction is 
\begin{equation}
\sigma_{\mu}^2=\sum_{ij}\bar r_i\bar r_jC_{ij}=\sum_{ij}\bar r_i\bar
r_j(r_ir_jV_{\mathrm{ext}}+\delta_{ij}r_i)>V_{\mathrm{ext}}\sum_{ij}r_i^2r_j^2/|r|^2=V_{\mathrm{ext}}|r|^2
\end{equation}

Hence, $\sigma_{\mu}^2/\sigma_{all}^2> \frac{V_{\mathrm{ext}}|r|^2}{|r|^2V_{\mathrm{ext}}+N\langle {r}
\rangle}$, which depends only weakly on $\langle {r} \rangle$, if $N\langle {%
r} \rangle$ is not much bigger than $V_{\mathrm{ext}}|r|^2\geq V_{\mathrm{ext}} N\langle {r} \rangle^2$%
, that is once again if noise correlations are not too small.

\subsubsection{Additional test for a common origin of shared variability}
\label{app:gainFlucTest} If rates are only measured up to an unknown offset $a$, the variance $\sigma_{\mu}^2/\sigma_{all}^2$, which is obtained from a projection on the apparent average response $r-a$ may not
be constant across stimuli. In this case, the variation of this quantity across stimuli is not a reliable indicator to exclude a gain fluctuation model as the source of observed correlations.
However, if a large part of the covariances can
be explained by a single component, it should be possible to reconstruct the
common origin from the stimulus dependent covariance matrices. From 
\begin{equation}
C(s)=\bigl(r(s)+a\bigr)^T\bigl(r(s)+a\bigr)V_{\mathrm{ext}}+D\bigl(r(s)+a\bigr)
\end{equation}
the vectors $v(s)=r(s)+a$ can be obtained approximately by finding the
eigenvector with the largest eigenvalue of $C$, by neglecting the contribution of $D\bigl(r(s)+a\bigr)$. This approximation can also be avoided by applying a factor analysis with a single latent component. Because $a$ is constant across stimuli, up to measurement errors of the estimated
covariances, the straight lines through the mean responses $r+x\bar v$, for $%
x\in \mathrm{I\!R}$ and normalized directions $\bar v$, will intersect in
the point $-a$. The best intersection point of multiple lines in the least
square sense is given by 
\begin{equation}
-\hat a=\Bigl(\sum_s\mathbb{I}-\bar v(s)\bar v(s)^T\Bigr)^{-1}\Bigl(\sum_s(%
\mathbb{I}-\bar v(s)\bar v(s)^T)r(s)\Bigr)
\end{equation}

and can be used to correct the average responses to $r^{\prime
}(s)=r(s)+\hat a$. We found that the analysis of $\sigma^2_{\mu}$ and $\sigma^2_{d}$ based
on the corrected responses lead to no qualitative change in the stimulus dependence,
indicating that a potential shared component is too weak to be identified based on
the available data. 

\subsubsection{Feed-forward model}
\label{app:FFmodel} 
Here we derive characteristic relations for the response statistics resulting in a feed-forward network. They  illustrate qualitative differences between the predictions of different models and will be used to extract model parameters from the data.
In the feed-forward model the mean responses and covariances are, from Eqs.\ (\ref{FFNrates}) and (\ref%
{covsFFN}) and allowing an offset in observed rates, 
\begin{equation}
r=Fr_{\mathrm{ext}},
\end{equation}
\begin{equation}
C_{ij}=\delta_{ij}(r_{i}+a)+\sum_kF_{ik}F_{jk}V_{\mathrm{ext},k}.
\end{equation}

In contrast to the recurrent network model (see below), the average covariance is not correlated to  the population averaged rate $\langle r \rangle$ across stimuli, Eq.\ (\ref{eq:covScalingFF}).  This follows after taking the average across neurons:
\begin{equation}  \label{eq:covScFFapp}
\langle {C_{ij}} \rangle_{i \neq j}=\frac{1}{N(N-1)}\sum_{i \neq
j,k}F_{ik}F_{jk}V_{\mathrm{ext},k}\approx N\langle {F_{ik}} \rangle_{ik}\langle {F_{jk}%
} \rangle_{jk}\langle {V_{\mathrm{ext},k}} \rangle_k=N\langle {F} \rangle^2\langle {V_{\mathrm{ext}}%
} \rangle.
\end{equation}
Input variances $V_{\mathrm{ext}}$ are assumed to be independent of the network structure and $%
F_{ik}$ independent of $F_{jk}$, which means that the strengths of connection of
an external neuron to different internal neurons are independent.

If $\langle {C_{ij}} \rangle_{i \neq j}$ is to be uncorrelated to $\langle r \rangle$  across stimuli,  the average input variance $\langle {V_{\mathrm{ext}}} \rangle$ needs to be approximately uncorrelated to  $%
\langle {r} \rangle$. This holds in our model even though $r_k=\sum_iF_{ik}r_{\mathrm{ext},k}$ and in each {\it input} channel the variance equals the strength of the input, $V_{\mathrm{ext},k}=|r_{\mathrm{ext},k}|$, if either the
distribution of inputs across neurons is approximately symmetric around 0,
or the distribution of column sums $\sum_iF_{ik}$ of the feed-forward matrix
(across columns) is symmetric around 0. The first case can be realized in a feed-forward network with a similar number of excitatory and inhibitory input channels. Intuitively, inhibitory inputs decrease the average output, but contribute positively to the average variance, and thus decorrelate the two quantities.

Formally, one compares the random
variable $\sum_kr_k=\sum_{ik}F_{ik}r_{\mathrm{ext},k}=\sum_kr_{\mathrm{ext},k}\sum_iF_{ik}$ to the
variable $\sum_kV_{\mathrm{ext},k}=\sum_k|r_{\mathrm{ext},k}|$. A random variable $x$ is
uncorrelated to its absolute value $|x|$ if its distribution is symmetric around 0. Here, the relevant variables are the elements of the input vector $r_{\mathrm{ext}}$, and their distribution is approximately symmetric around 0, if the variance $\var{r_{\mathrm{ext}}}$ is much larger than their mean $\langle r_{\mathrm{ext}} \rangle$, that is if $\rho_E = \var{r_{\mathrm{ext}}}/\langle r_{\mathrm{ext}} \rangle \gg 1$. Consequently, if the variance of external inputs across stimuli is high, the population averaged covariances in a feed-forward network are uncorrelated to the population response.

In the following, we motivate that the stimulus dependence of the variances projected along different directions in the feed-forward model is different from the one in the gain fluctuation model. The sum of the variances is 
\begin{equation}
\sigma_{all}^2=N(\langle {r} \rangle+a)+\sum_{i,k}F_{ik}^2V_{\mathrm{ext},k}
\approx N\bigl(\langle {r} \rangle+a\bigr)+N^2\langle {V_{\mathrm{ext}}} \rangle\langle {F^2} \rangle.
\label{eq:sigAllFF}
\end{equation}

Due to the Poisson spike generation, there is a linear
contribution in $\langle {r} \rangle$ to the average variance, $\langle {%
C_{ii}} \rangle_{i}=\sigma_{all}^2/N$. This variability does not contribute
to covariances, because spikes are generated independently across neurons,
in contrast to the recurrent model, where the spiking or not spiking of a
neuron directly influences post-synaptic firing rates.

The variance projected onto the diagonal direction is 
\begin{equation}
\sigma_{d}^2=\langle {r} \rangle+a+\frac{1}{N}\sum_{ijk}F_{ik}F_{jk}V_{\mathrm{ext},k}
\approx \langle {r} \rangle+a+N^2\langle {V_{\mathrm{ext}}} \rangle\langle {F} \rangle^2.
\end{equation}
Consequently, the ratio $\sigma_{d}^2/\sigma_{all}^2$ depends only weakly on $%
\langle {r} \rangle$ if the term $N\langle V_{\mathrm{ext}}\rangle\langle {F} \rangle^2$ is large
against $\langle {r} \rangle+a$, that is if the sum of covariances is larger
than the sum of variances. This is the case, if correlation coefficients
are larger than of the order $O(1/N)$. The variance projected onto the
direction of the mean response is 
\begin{equation}
\sigma_{\mu}^2=\sum_{i}(r_i+a)\bar r_i^2+\sum_{ijk}F_{ik}F_{jk}V_{\mathrm{ext},k}\bar
r_i\bar r_j\approx a+\frac{\sum_ir_i^3}{|r|^2}+\frac{\langle {r} \rangle^2}{%
|r|^2}N^3\langle {F} \rangle^2\langle {V_{\mathrm{ext}}} \rangle.
\end{equation}
To see that the ratio $\sigma_{\mu}^2/\sigma_{all}^2$ strongly depends on
the population response, in contrast to $\sigma_{d}^2/\sigma_{all}^2$, assume again that the sum across variances in the
first term $a+\frac{\sum_ir_i^3}{|r|^2}$ is not too large against the sum
across covariances in the second term. The second term depends strongly on $\langle r \rangle$: it is 0 for $\cos(r,d)=0$%
. For $\cos(r,d)=1$ it becomes $N^2\langle {F} \rangle^2\langle {V_{\mathrm{ext}}}
\rangle $. In this case, $\sigma_{\mu}^2/\sigma_{all}^2$ is not too small, if $%
(\langle {r} \rangle+a)/(NV_{\mathrm{ext}})$ is not much larger than 1 (assuming that $%
\langle {F^2} \rangle/\langle {F} \rangle^2$ is of order one).

Finally, we calculate the average noise correlation coefficient (across
stimuli and neuron pairs) 
\begin{equation}
c_N=\langle {\frac{C_{ij}(s)}{\sqrt{C_{ii}(s)C_{jj}(s)}}} \rangle_{s, i\neq
j}.
\end{equation}
For a given stimulus, we assume that the $C_{ij}$ are approximately
independent, so that we can write 
\begin{equation}
\langle {\frac{C_{ij}(s)}{\sqrt{C_{ii}(s)C_{jj}(s)}}} \rangle_{i\neq
j}\approx\frac{\langle {C_{ij}(s)} \rangle_{i\neq j}}{\langle {C_{ii}(s)}
\rangle_{i}}.
\end{equation}
Then, from Eqs.\ (\ref{eq:covScFFapp}) and (\ref{eq:sigAllFF}), 
\begin{equation}  \label{eq:cNFFapp}
c_N=\frac{N\langle {F} \rangle^2\langle {V_{\mathrm{ext}}} \rangle}{\langle {r}
\rangle+a+N\langle {V_{\mathrm{ext}}} \rangle\langle {F^2} \rangle}.
\end{equation}
Signal correlations can be calculated analogously as for the recurrent
model, see below.

\subsubsection{Recurrent network model}
\label{app:recurrent} The response distributions in the recurrent network are characterized by
the average and covariances of
responses  given in Eqs.\ (\ref{HawkesRates}) and (\ref{HawkesCovs}), 
\begin{equation}
r=Br_{\mathrm{ext}}
\end{equation}
and 
\begin{equation}
C=BD[r_{\mathrm{eff}}]B,
\end{equation}
with $r_{\mathrm{eff}}=r+a+V_{\mathrm{ext}}$, including an offset $a$. To compare this model to the alternative scenarios described in the previous sections, we calculate the relation between average (co)variances and the projected variances and  $r$ as well as the signal and noise correlations. 

We first derive Eqs.\ (\ref{eq:varScalingRec}) and (\ref{eq:covScalingRec}). The average covariance is 
\begin{equation}
\langle {C_{ij}} \rangle_{i \neq j}=\frac{1}{N(N-1)}\sum_{i\neq
j,k}B_{ik}B_{jk}r_{\mathrm{eff},k}\approx N \langle {B_{ik}B_{jk}} \rangle_{ijk}%
\langle {r_{\mathrm{eff},k}} \rangle_k.
\end{equation}
We assume that $N$ is sufficiently large and that the external input characterized by $V_{\mathrm{ext}}, r_{\mathrm{ext}}$
does not depend on the recurrent network. In that case $r_{\mathrm{eff},k}$ is
approximately uncorrelated to $B_{ik}$ (including the contribution of $r_k$, because $\langle {B_{ij}r_k}
\rangle=\langle {B_{ij}\sum_lB_{kl}r_{\mathrm{ext},l}} \rangle\approx\langle {B_{ij}}
\rangle\langle {\sum_lB_{kl}r_{\mathrm{ext},l}} \rangle$). The term $\langle {r_{\mathrm{eff}}}
\rangle=\langle {r} \rangle+a+\langle {V_{\mathrm{ext}}} \rangle$ is linear in $\langle {r%
} \rangle$, with an uncorrelated contribution $\langle {V_{\mathrm{ext}}} \rangle$, see
Appendix \ref{app:FFmodel}. If the elements of  $B$ are pairwise independent, 
\begin{equation}  \label{eq:covScalingRecApp}
\langle {C_{ij}} \rangle_{i \neq j}=N\langle {B} \rangle^2(\langle {r}
\rangle+a+\langle {V_{\mathrm{ext}}} \rangle).
\end{equation}

Under the same assumptions, the sum of variances is 
\begin{align}
\begin{split}
\sigma_{all}^2&=\sum_{i,k}B_{ik}^2r_{\mathrm{eff},k}=\sum_{ik}B_{ik}^2(V_{\mathrm{ext},k}+r_k+a)
\\
&\approx N\langle {B^2} \rangle\sum_k(V_{\mathrm{ext},k}+r_k+a) = N^2 \langle {B^2}
\rangle(\langle {r} \rangle+a+\langle {V_{\mathrm{ext}}} \rangle).  \label{eq:sigAllRec}
\end{split}%
\end{align}
Using the same argument as in the feed-forward scenario following Eq.\ (\ref%
{eq:covScFFapp}), the term $N^2\langle {B^2} \rangle\langle {V_{\mathrm{ext}}} \rangle$
only contributes with an offset to the relation between $\langle {r} \rangle$
and $\langle {C_{ii}} \rangle_{i}=\sigma_{all}/N$. 

Next, we show that the projected variances in this scenario have a similar dependence on $\langle r \rangle$ as the ones in the feed-forward model.
The projected variance on the diagonal is 
\begin{equation}
\sigma_{d}^2=\frac{1}{N}\sum_{ij,k}B_{ik}B_{jk}r_{\mathrm{eff},k}\approx N^2\langle {B%
} \rangle^2(\langle {r} \rangle+a+\langle {V_{\mathrm{ext}}} \rangle).
\end{equation}
Consequently, in the ratio ${\sigma_{d}^2}/{\sigma_{all}^2}$ the dependence on $%
\langle {r} \rangle$ is canceled out.

The projection along the mean response direction, 
\begin{equation}
\sigma_{\mu}^2=\sum_{ij,k}B_{ik}B_{jk}\Bigl(r_k+a+V_{\mathrm{ext},k}\Bigr)\bar r_i\bar
r_j\approx N^3\frac{\langle {r} \rangle^2}{|r|^2}\langle {B}
\rangle^2(\langle {r} \rangle+a+\langle {V_{\mathrm{ext}}} \rangle),
\end{equation}
depends strongly on $\langle {r} \rangle$. Because $\cos(d,r)=0$ implies $%
\langle {r} \rangle=0$, $\sigma_{\mu}^2/\sigma_{all}^2=0$ in this case, and
for $\cos(d,r)=1$, with $|r|^2=N\langle {r} \rangle^2$ one gets $%
\sigma_{\mu}^2/\sigma_{all}^2=\langle {B} \rangle^2/\langle {B^2} \rangle$.

Apart from the raw covariances, we are interested the average noise
correlation coefficient (across stimuli and neuron pairs).
As for the feed-forward model, we approximate 
$c_N 
\approx \frac{\langle {C_{ij}} \rangle_{s,i\neq j}}{\langle {C_{ii}(s)}
\rangle_{s,i}}
$. 
 From Eqs.\ (\ref{eq:covScalingRecApp}) and (\ref%
{eq:sigAllRec}) 
\begin{equation}
c_N=\frac{N\langle {B} \rangle^2(\langle {r} \rangle+a+\langle {V_{\mathrm{ext}}} \rangle)%
}{N\langle {B^2} \rangle(\langle {r} \rangle+a+\langle {V_{\mathrm{ext}}} \rangle)}=\frac{%
\langle {B} \rangle^2}{\langle {B^2} \rangle},
\end{equation}
and because $\langle {B^2} \rangle=\mathrm{var}({B})+\langle {B} \rangle^2$ 
\begin{equation}  \label{Eq:noiseCorr}
c_N=\frac{1}{1+\mathrm{var}({B})/\langle {B} \rangle^2}=\frac{1}{1+\rho}.
\end{equation}
We see  that the correlation coefficient depends on the relative
variability of the network elements. The variances, resulting
from the variances of the input channels, are determined by the mean of the
square elements of the network. By contrast, the covariances depend on the
effective weights of inputs to the neuron pairs, and hence on the square of
the mean weight. Correspondingly, we consider the signal covariances $C^S_{ij}(s)=%
\mathrm{cov}({r_i(s),r_j(s)})_s$, 
\begin{equation}
C^S_{ij}=\mathrm{cov}({\sum_{kl}B_{ik}r_{\mathrm{ext},k}(s),B_{jl}r_{\mathrm{ext},l}(s)}%
)_s=\sum_{kl}\mathrm{cov}({B_{ik}r_{\mathrm{ext},k}(s),B_{jl}r_{\mathrm{ext},l}(s)})_s.
\end{equation}
For $i\neq j$, one gets 
\begin{align}
\begin{split}
C^S_{ij}&=\sum_{kl}B_{ik}B_{jk}\mathrm{cov}({r_{\mathrm{ext},k}(s),r_{\mathrm{ext},l}(s)})_s \\
&=\sum_k B_{ik}B_{jk}\mathrm{var}({r_{\mathrm{ext},k}(s)})_s+\sum_{k\neq l}B_{ik}B_{jk}%
\mathrm{cov}({r_{\mathrm{ext},k}(s),r_{\mathrm{ext},l}(s)})_s.
\end{split}%
\end{align}
Averaged across neurons 
\begin{equation}
\langle {C^S_{ij}} \rangle_{i\neq j}=N\langle {B} \rangle^2\mathrm{var}({r_{\mathrm{ext}}}%
)+N(N-1)\langle {B} \rangle^2c_{\mathrm{in}}\mathrm{var}({r_{\mathrm{ext}}}).
\end{equation}
For the signal variances, $i=j$, which correspond to the variance of the rates across stimuli, 
\begin{align}
\begin{split}
\langle {C^S_{ii}(s)} \rangle_{s}&=\sum_k\mathrm{var}({B_{ik}r_{\mathrm{ext},k}}%
)+\sum_{k\neq l}\mathrm{cov}({B_{ik}r_{\mathrm{ext},k},B_{il}r_{\mathrm{ext},l}}) \\
&=\sum_kB_{ik}^2\mathrm{var}({r_{\mathrm{ext},k}})+\sum_{k\neq l}\mathrm{cov}({%
r_{\mathrm{ext},k},r_{\mathrm{ext},l}})B_{ik}B_{il}. \\
\end{split}%
\end{align}
Averaged across neurons, with $\langle {B^2} \rangle=\langle {B} \rangle^2+%
\mathrm{var}({B})$: 
\begin{equation}
\langle {C^S_{ii}(s)} \rangle_{s,i}=N\mathrm{var}({r_{\mathrm{ext}}})(\langle {B}
\rangle^2+\mathrm{var}({B}))+N(N-1)\langle {B} \rangle^2\mathrm{var}({r_{\mathrm{ext}}}%
)c_{\mathrm{in}}  \label{eq:recRateVariance}.
\end{equation}

Their ratio is 
\begin{align}
\begin{split}
\frac{\langle {C_{ij}} \rangle }{\langle {C_{ii}} \rangle}&=\frac{N\langle {B}
\rangle^2\mathrm{var}({r_{\mathrm{ext}}})(1+(N-1)c_{\mathrm{in}})}{N\mathrm{var}({r_{\mathrm{ext}}})(\langle {B}
\rangle^2+\mathrm{var}({B}))+N(N-1)\langle {B} \rangle^2\mathrm{var}({r_{\mathrm{ext}}}%
)c_{\mathrm{in}}}\\
&=\frac{1+(N-1)c_{\mathrm{in}}}{1+\mathrm{var}({B})/\langle {B}
\rangle^2+(N-1)c_{\mathrm{in}}}.
\end{split}%
\end{align}

This results in an approximate expression for the average signal correlation
coefficient 
\begin{equation}  \label{Eq:signalCorr}
c_S=\langle {\frac{C^S_{ij}}{\sqrt{C_{ii}^SC_{jj}^S}} } \rangle_{i,j}\approx\frac{\langle {C^S_{ij}} \rangle }{\langle {C^S_{ii}} \rangle}=\frac{%
1+(N-1)c_{\mathrm{in}}}{1+\mathrm{var}({B})/\langle {B} \rangle^2+(N-1)c_{\mathrm{in}}}
.
\end{equation}

\subsubsection{Estimation of network model parameters}

\label{app:estimation} We extract parameters of the recurrent and feed-forward network models from the data, both to test the consistency of the models 
with the data and to interpret the observed variability.

Because rates and covariance matrices were measured for many different stimuli and thus provide a large number of constraints, one approach would be to infer as much information as possible about the full connectivity matrices $B$ or $F$ from the data.
However, due to the relatively small number of trials for each stimulus, we use a model with few parameters. The set of parameters  consists of the  network parameters $\langle 
{B} \rangle$ and $\mathrm{var}({B})$ ($\langle {F} \rangle$ and $\mathrm{var}%
({F})$, respectively) as well as the parameters of the input ensemble, $%
\langle {r_{\mathrm{ext}}} \rangle$, $\mathrm{var}({r_{\mathrm{ext}}})$ and $c_{\mathrm{in}}$. 

In particular, we want to infer the ratios $\rho_E={\mathrm{var}({r_{\mathrm{ext}}})}/ {\langle {r_{\mathrm{ext}}} \rangle^2}$ and $%
\rho={\mathrm{var}({B})}/{\langle {B} \rangle^2}$ (and correspondingly for $F$). Based on these, we can generate surrogate data to test if the observed scaling of average covariances with average rates is more consistent with the recurrent or the feed-forward model, Eq.\ (\ref{eq:covScalingRec}) or (\ref{eq:covScalingFF}). The experimental data provides constraints in the form of the population and stimulus averaged rates, their variances and the noise and signal correlation coefficients.
 In the models, rates
are given by Eqs.\ (\ref{HawkesRates}) and (\ref{FFNrates}), respectively.
The population averaged mean response thus is 
\begin{equation}  \label{eq:meanRate}
\langle {r} \rangle=N\langle {B} \rangle\langle {r_{\mathrm{ext}}} \rangle \ \text{or} \
\langle {r} \rangle=N\langle {F} \rangle\langle {r_{\mathrm{ext}}} \rangle.
\end{equation}

The variance of rates across stimuli, see Eq.\ (\ref{eq:recRateVariance}),
is 
\begin{equation}  \label{eq:varRate}
\mathrm{var}({r})=N\mathrm{var}({r_{\mathrm{ext}}})(\langle {B} \rangle^2+\mathrm{var}({B}%
)+(N-1)\langle {B} \rangle^2c_{\mathrm{in}})
\end{equation}
such that the relative variance of rates 
\begin{equation}  \label{eq:rateRatio}
\frac{\mathrm{var}({r})}{\langle {r} \rangle^2}=\frac{\mathrm{var}({r_{\mathrm{ext}}})}{%
\langle {r_{\mathrm{ext}}} \rangle^2}\frac{1+\rho+(N-1)c_{\mathrm{in}}}{N}
\end{equation}
depends on the input signal to noise ratio $\rho_E=\frac{\mathrm{var}({r_{\mathrm{ext}}})%
}{\langle {r_{\mathrm{ext}}} \rangle^2}$ and the variability of the network elements, $%
\rho=\frac{\mathrm{var}({B})}{\langle {B} \rangle^2}$, or $\rho=\frac{%
\mathrm{var}({F})}{\langle {F} \rangle^2}$ respectively.

The estimates for the remaining parameters $\rho$ and $c_{\mathrm{in}}$ are obtained from the
measured values of the ratio of covariances to variances, which correspond
approximately to the average coefficients of noise and signal correlations, 
\begin{equation}
\langle {\frac{\langle {C_{ij}} \rangle_{i\neq j }}{\langle {C_{ii}}
\rangle_{i}}} \rangle_s\approx c_N,
\end{equation}
and 
\begin{equation}
\frac{\langle {\mathrm{cov}({r_i,r_j})} \rangle_{i\neq j}}{\langle {\mathrm{%
var}({r_i})} \rangle_i}\approx c_S,
\end{equation}

using Eqs.\ (\ref{Eq:noiseCorr}) and (\ref{Eq:signalCorr}). Together with Eq.\ (\ref{eq:rateRatio}), these relations provide the necessary constraints for the network models.
Strictly
speaking, these equations are valid only for the recurrent network, while
for the feed-forward model, Eq.\ (\ref{eq:cNFFapp}) is relevant. In this
case $\rho$ is overestimated, which results in a lower bound for $\rho_E$,
and thus a conservative estimate of the input variability.

Under additional assumptions, we can also choose the absolute values of the
parameters, for example $\langle {r_{\mathrm{ext}}} \rangle$ and $\langle {F} \rangle$,
such that mean rates and mean covariances correspond to experimental ones.
The mean rates (\ref{eq:meanRate}) constrain the product of the two
parameters 
\begin{equation}
\langle {r} \rangle=N\langle {F} \rangle\langle {r_{\mathrm{ext}}} \rangle.
\end{equation}

From Eq.\ (\ref{eq:covScFFapp}) it follows that 
\begin{equation}
\langle {C_{ij}} \rangle_{i\neq j }=N\langle {F} \rangle^2\langle {V_{\mathrm{ext}}}
\rangle,
\end{equation}
and we assume that $V_{\mathrm{ext}}=|r_{\mathrm{ext}}|$ to relate mean input and input variance. The
distribution of $V_{\mathrm{ext}}$ thus is a folded normal distribution, with 
\begin{equation}
\langle {|r_{\mathrm{ext}}|} \rangle=\sqrt{\mathrm{var}({r_{\mathrm{ext}}})}\sqrt{2/\pi}%
e^{-1/2\rho_x}+\langle {r_{\mathrm{ext}}} \rangle\Bigl(1-2\Phi(-\sqrt{1/\rho_x})\Bigr)
\end{equation}
(with the cumulative normal $\Phi$ ), and from this one finds 
\begin{equation}
\langle {C_{ij}} \rangle=\langle {F} \rangle\langle {r} \rangle\Bigl[\sqrt{%
\rho_x}e^{-1/2\rho_x}+1-2\Phi(-\sqrt{1/\rho_x}) \Bigr].
\end{equation}
To set the absolute values of input versus network strength we made
assumptions regarding the relation between input strength and input variance
across trials. If we measure the strength of the dependence between mean
response and mean covariance, Eq.\ (\ref{eq:covScFFapp}) by the ratio
intercept/slope of a linear fit, however, the result is independent of the
absolute values of $\langle {F} \rangle$ and $\langle {r_{\mathrm{ext}}} \rangle$, as
well as a potential linear factor relating $V_{\mathrm{ext}}$ and $|r_{\mathrm{ext}}|$.

\subsection{Covariances and rates on a population level in regular networks}
\label{app:popModel} 
Here we derive the macroscopic Eqs.\ (\ref{eq:popRates}) and (\ref{eq:popCovs}) for the responses of the two populations: the average response of population $I$, $R_I\equiv\sum_{i \in I}r_i$, and the covariances between responses of populations $L,K$, $\Sigma_{KL}\equiv\sum_{k\in K,l\in L}C_{kl}$, in the regular networks defined in Section \ref{sec:twopop} depend only on the coupling between populations, see below.

The two populations are of size $N$, and weights between connected neurons are of equal strength.  
By definition, the regularity of the network connectivity implies that for any neuron $l^{\prime }\in$ from population $L$, the sum $\sum_{k\in
K}G_{kl^{\prime }}$ is identical, and we can define the population coupling matrix, $\Gamma$, by
\begin{equation}
\Gamma_{KL}\equiv\sum_{k\in K}G_{kl^{\prime }}= \frac{1}{N}\sum_{k\in K,l\in
L}G_{kl}=n_{KL}g,
\end{equation}
where $g$ is the weight of the connections. We want to express $R$ and $\Sigma$ in terms of $\Gamma$.
On the neuron level, rates, $r$, and covariances, $C$, depend on the transfer matrix, $B=(\id-G)^{-1}$. We will define an analogous population transfer matrix $\mathrm{P}$ that depends only on  $\Gamma$ and  show that $R$ and $\Sigma$ can be written in terms of $\mathrm{P}$. We define
\begin{equation}
\mathrm{P}\equiv(\mathbb{I}-\Gamma)^{-1}=\mathbb{I}+\Gamma+\Gamma^2+\dots=\sum_{m=0}^{\infty}\Gamma^m.
\end{equation}
To relate $\mathrm{P}$ to the microscopic quantities $r$ and $C$, we will use that
\begin{equation}  \label{populationTransfer}
\mathrm{P}_{KL}= \frac{1}{N}\sum_{k\in K,l\in L}B_{kl}=\sum_{k\in K}B_{kl^{\prime }}
\end{equation}
for any $l'\in L$, in complete analogy to $\Gamma$.
 To see this, we note the corresponding relation for all of the individual terms $m$ 
\begin{equation}
[\Gamma^m]_{KL}=\frac{1}{N}\sum_{k\in K,l\in L}[G^m]_{kl}.
\end{equation}
As an example consider $m=2$. Using the regularity, we see that 
\begin{equation}
\sum_{k\in K,l\in L}[G^2]_{kl}=\sum_{l\in L,k\in K,i\in
I,I}G_{ki}G_{il}=\sum_{l\in L,i\in
I,I}\Gamma_{KI}G_{il}=N\sum_I\Gamma_{KI}\Gamma_{IL}=[\Gamma^2]_{KL}N.
\end{equation}
For the rate of population $I$,  $R_I$, we then find with (\ref{populationTransfer}) that
\begin{equation}
R_I\equiv\sum_{i \in I}r_i=\sum_{i\in I}\sum_K\sum_{k\in
K}B_{ik}r_{\mathrm{ext},k}=\sum_K\sum_{i\in I,k\in
K}B_{ik}r_{\mathrm{ext},k}=\sum_K\mathrm{P}_{IK}NR_{\mathrm{ext},K},
\end{equation}
because $r_{\mathrm{ext},k}=R_{\mathrm{ext},K}$ for any $k\in K$, and thus 
\begin{equation}
R=\mathrm{P}NR_{\mathrm{ext}}=(\mathbb{I}-\Gamma)^{-1}NR_{\mathrm{ext}},
\end{equation}
which is Eq.\ (\ref{eq:popRates}).
Similarly, for the covariances of the population responses, $\Sigma$:
\begin{equation}
\Sigma_{KL}\equiv\sum_{k\in K,l\in L}C_{kl}=\sum_I\sum_{k\in K,l\in L,i\in
I}B_{ki}B_{li}r_i=\sum_{l\in L}\sum_{I}\sum_{i\in I}\mathrm{P}_{KI}B_{li}r_i
\end{equation}
\begin{equation}
=\sum_{I,i\in I}\mathrm{P}_{KI}\mathrm{P}_{LI}r_i=\sum_I\mathrm{P}_{KI}\mathrm{P}_{LI}R_I, 
\end{equation}
which corresponds to Eq.\ (\ref{eq:popCovs}), 
\begin{equation}
\Sigma=(\mathbb{I}-\Gamma)^{-1}D[R](\mathbb{I}-\Gamma)^{-1},
\end{equation}
(with the diagonal matrix $D[R]$ with $D[R]_{IJ}=\delta_{IJ}R_I$).

\subsubsection{Effect of cross-coupling on correlations on a population level%
}
In this section, we evaluate how noise and signal
correlations in the population activity arise from direct and effective coupling.
\label{app:popCorrs} The strength of the cross-coupling between populations, $%
\Gamma_c$, determines via the effective cross-coupling, $P_c$, of the
population transfer matrix, 
\begin{equation}  \label{eq:Bpop}
\mathrm{P}=(\mathbb{I} -\Gamma)^{-1}=\frac{1}{(1-\Gamma_s)^2-\Gamma_c^2} 
\begin{pmatrix}
1-\Gamma_s & \Gamma_c \\ 
\Gamma_c & 1-\Gamma_s \\ 
\end{pmatrix}
\equiv%
\begin{pmatrix}
P_s & P_c \\ 
P_c & P_s%
\end{pmatrix},
\end{equation}
 the population rates and covariances. The network is unstable if the
eigenvalues of $\Gamma$ are larger than one, which corresponds to the
constraint $\Gamma_s+\Gamma_c<1$, and therefore $P_c<P_s$.

On the population level, the effects of cross-coupling between differently
tuned neurons on correlations can be expressed in terms of the variance and
the mean of the elements of $\mathrm{P}$, similar as for the random effective network
model. To see this, we consider an ensemble of (two-component) input vectors 
$R_{\mathrm{ext}}(s)=\bigl(r_{\mathrm{ext}}(s),r_{\mathrm{ext}^{\prime }}(s)\bigr)^T$ with external input
components $r_{\mathrm{ext}},r_{\mathrm{ext}^{\prime }}$ chosen with mean and variance $%
\mu_{r_{\mathrm{ext}}},\sigma^2_{r_{\mathrm{ext}}}$, independently across stimuli $s$.

The signal covariance matrix of mean responses $R=\mathrm{P}R_{\mathrm{ext}}$ across the
stimulus ensemble is $\Sigma^S_{ij}=\mathrm{cov}({R_i,R_j})$ which is 
\begin{equation}
\Sigma^S=\sigma_{r_{\mathrm{ext}}}^2%
\begin{pmatrix}
P_s^2+P_c^2 & 2P_sP_c \\ 
2P_sP_c & P_s^2+P_c^2%
\end{pmatrix}%
\end{equation}

The average signal correlation for the population activity is 
\begin{equation}
\begin{split}
\langle {c_S} \rangle&=\frac{1}{4}\sum_{ij}\frac{\Sigma^S_{ij}}{\sqrt{%
\Sigma^S_{ii}\Sigma^S_{jj}}}=1+\frac{2P_sP_c}{P_s^2+P_c^2}=\frac{(P_s+P_c)^2%
}{(P_s+P_c)^2+(P_s-P_c)^2} \\
&=\frac{\langle P \rangle^2}{\langle P \rangle^2+\mathrm{var}(P)}=\frac{\langle {P} \rangle^2}{\langle {%
P^2} \rangle}=\frac{1}{1+\mathrm{var}(P)/\langle P \rangle^2}
\end{split},
\end{equation}
using $\langle P \rangle=(P_s+P_c)/2$ and $%
2\mathrm{var}(P)=(P_s-(P_s+P_c)/2)^2+(P_c-(P_s+P_c)/2)^2=(P_s-P_c)^2/2$. The
normalized variance of the elements of $\mathrm{P}$ thus determines the strength of signal
correlations, and one can interpret $\rho=\mathrm{var}(P)/\langle P \rangle^2$ as a measure
for coupling strength.

A similar relation holds for the average noise correlations, if we average  first
 across stimuli, then across neurons: the average of the noise covariance matrix across stimuli is $\langle {\Sigma(s)}
\rangle_s=\langle {PR(s)P^T} \rangle=P\langle {R} \rangle P^T$. Because $\langle {%
R_{\mathrm{ext}}} \rangle=\langle {R_{\mathrm{ext}}^{\prime }} \rangle$, the average rates of both populations are identical, and the average across population pairs can be evaluated just as for the signal correlations.

\subsubsection{Condition for favorable correlations}
Here,  we show that strong connections between populations can induce correlations
that are beneficial for stimulus discrimination, when compared to shuffled trials.
Shuffling correlations is favorable for stimulus discrimination if the
covariances within each populations are stronger than the ones across
populations. To see this, assume that the responses of the excitatory
populations for two stimuli are $R(s_1)=(R_0+\Delta R,R_0)$ and $%
R(s_2)=(R_0,R_0+\Delta R)$. We also assume that
the population covariance matrix is stimulus independent, $%
\Sigma(s_1)=\Sigma(s_2)=%
\begin{pmatrix}
\Sigma_{EE} & \Sigma_{E^{\prime }E} \\ 
\Sigma_{E^{\prime }E} & \Sigma_{E^{\prime }E^{\prime }} \\ 
\end{pmatrix}%
$ and that $\Sigma_{EE}= \Sigma_{E^{\prime }E^{\prime }}$. Calculating the
most discriminative direction $w$ (using Eq.\ (\ref{eq:mostDD})  on the macroscopic variables) yields an eigenvector of $\Sigma$ ,  $(1,-1)$, and
the variance in that direction corresponds to the smaller eigenvalue of $%
\Sigma$, 
\begin{equation*}
\sigma_w^2=\Sigma_{EE}-\Sigma_{E^{\prime }E}=\sum_{i \in
E}C_{ii}+\sum_{i\neq j \in E}C_{ij}-\sum_{i\in E, k\in E^{\prime }}C_{ik}.
\end{equation*}
Covariances can be considered helpful, if this eigenvalue is smaller when
compared to a shuffled version where all covariances are set to 0, such that $%
\sigma_{w^{\prime }}^2=\sum_{i\in E}C_{ii}$. This is exactly the case if the sum of covariances across populations, $\sum_{i\in E, k \in E'}C_{ik}$, is larger than the one within a population, $\sum_{i\neq j \in E}C_{ij}$.
Although $\Sigma_{EE}\geq
\Sigma_{E^{\prime }E}$, it is possible that $\langle {C_{ij}}
\rangle_{i\neq j\in E}<\langle {C_{ij}} \rangle_{i\in E,j\in E^{\prime }}$,
because of the contribution of the variances $C_{ii}$ to $\Sigma_{EE}$.

We want to find a condition for which such a removal of covariances by  shuffling
 decreases noise in the relevant
direction. For simplicity, let us assume that $R_{\mathrm{ext}}=R_{\mathrm{ext}^{\prime }}\equiv R$.
Then 
\begin{equation}
\Sigma=R%
\begin{pmatrix}
P_s^2+P_c^2 & 2P_sP_c \\ 
2P_sP_c & P_s^2+P_c^2%
\end{pmatrix}%
\end{equation}
The average pairwise across-covariance is $C_{across}=\Sigma_{E^{\prime
}E}/N^2$. Approximately, the average neuron variance differs from the
average within-covariance only by the contribution of the rate: $\langle {%
C_{ii}} \rangle\approx R/N+C_{within}$. 
From this we get for the population
variance $\Sigma_{EE}=R(P_s^2+P_c^2)=N\langle {C_{ii}}
\rangle+N(N-1)C_{within}=R+N^2C_{within}$. For favorable correlations, we
need $C_{across}>C_{within}$, 
\begin{equation}
\frac{2P_sP_cR}{N^2}>\frac{R(P_s^2+P_c^2)-R}{N^2}
\end{equation}
and since $P_s>P_c$ 
\begin{equation}
1>(P_s-P_c)^2\Leftrightarrow 1>P_s-P_c
\end{equation}
Using Eq.\ (\ref{eq:Bpop}) for the entries of $P$, this means that 
\begin{equation}
1>\frac{1-\Gamma_s-\Gamma_c}{(1-\Gamma_s)^2-\Gamma_c^2}\Leftrightarrow
\Gamma_s(1-\Gamma_s)>\Gamma_c(1-\Gamma_c).
\end{equation}
The function $\Gamma(1-\Gamma)$ is a parabola through the points (0,0) and
(0,1), with its minimum at (1/2,-1/4). We have $|\Gamma_s|,|\Gamma_c|<1$, so
both sides are $<0$ and $\Gamma_c<1-\Gamma_s$. Thus, the condition can be
fulfilled if $\Gamma_s<1/2$ and within-coupling is smaller than cross-coupling, $\Gamma_s<\Gamma_c<1-\Gamma_s$.

\subsubsection{Linear Fisher information}
\label{app:secLFI}
We use the signal-to-noise ratio, $S$, to measure how well discrete pairs of stimuli in a given ensemble can be discriminated, Section \ref{sec:discrimination}. In a model network, we have access to all possible input dimensions, and we can use an alternative measure that combines these possible stimulus dimensions. We will use this measure to compare the effects of different network scenarios on the representation of general stimuli.
 If a high-dimensional stimulus varies continuously, the
information in the response distribution about the stimulus can be
measured by the Fisher information matrix \cite{Abbott1999}. An approximate value, the
`linear Fisher information', is obtained by neglecting the information in
the stimulus dependence of the covariance matrix, and the entries of the
linear Fisher Information matrix are defined by 
\begin{equation}
I_{mn}=\partial_mR^T\Sigma^{-1}\partial_n R,
\end{equation}
for the population covariance matrix $\Sigma$ and population response $R$.
Here, $\partial_mR$ is the derivative of the response vector with respect to
stimulus coordinate $m$. 
Our stimulus dimensions are the coordinates of the input vector $R_{\mathrm{ext}}$ and output in a network model is $R=\mathrm{P}NR_{\mathrm{ext}}$. It is easy to see that $%
\partial_mR$ is given by the mth column of $\mathrm{P}$, and hence 
\begin{equation}
I=\mathrm{P}^{T}\Sigma^{-1}\mathrm{P}.
\end{equation}
Generally, if the input depends linearly on a one-dimensional stimulus $s$, $%
R_{\mathrm{ext}}(s)=R_{\mathrm{ext},0}+s\Delta R_{\mathrm{ext}}$, the information about changes in this direction
 is $\Delta R_{\mathrm{ext}}^TI\Delta R_{\mathrm{ext}}$. As a measure for the information about stimulus changes in all possible directions, we use the trace of $I$, $\mathrm{Tr}(I)=\sum_iI_{ii}$.

Covariances in a recurrent network with external noise are $%
\Sigma=\mathrm{P}(D[R]+\Sigma_{\mathrm{ext}})\mathrm{P}^{T}$, where $\Sigma_{\mathrm{ext}}$ is the diagonal $2\times 2$
matrix describing the variance of external input to the two populations.
This results in a recurrent linear Fisher Information matrix 
\begin{equation}  \label{app:LFIrec}
I^r=(D[R]+\Sigma_{\mathrm{ext}})^{-1}.
\end{equation}
This can be compared to a feed-forward scenario with identical transfer
matrix, $F=\mathrm{P}$, where $\Sigma=\mathrm{P}\Sigma_{\mathrm{ext}}\mathrm{P}^{T}+D[R]$, so that 
\begin{equation}  \label{app:LFIFF}
I^f=\mathrm{P}^{T}(\mathrm{P}\Sigma_{\mathrm{ext}}\mathrm{P}^{T}+D[R])^{-1}\mathrm{P}.
\end{equation}

\bibliographystyle{plos2015}

\end{document}